\documentclass[journal]{IEEEtran}
\IEEEoverridecommandlockouts
\usepackage{listings}
\usepackage{xcolor}
\usepackage{booktabs}
\usepackage{multirow}
\usepackage{url}

\lstdefinelanguage{ProVerif}{
  morekeywords={
    type,fun,reduc,equation,event,query,process,
    let,in,out,new,if,then,else,free,private
  },
  sensitive=true,
  morecomment=[l]{(*},
  morestring=[b]"
}

\usepackage{cite}
\usepackage{amsmath,amssymb,amsfonts}
\usepackage{algorithmic}
\usepackage{graphicx}
\usepackage{textcomp}
\usepackage{xcolor}
\usepackage{array}
\usepackage{float}
\usepackage{tikz}
\usetikzlibrary{positioning, shapes.multipart, arrows.meta}

\usepackage{amsthm}
\newtheorem{theorem}{Theorem}

\title{Quantum-Resilient, PUF-Based, and Anonymity-Preserving Authentication and Key Agreement Scheme for UAV Networks}

\author{Bhavik Yadav, Bhavya Dixit, Gaurav S. Kasbekar,~\IEEEmembership{Senior Member,~IEEE} 
\thanks{The authors are with the Department of Electrical Engineering, Indian Institute of Technology (IIT) Bombay, Mumbai, 400076, India. Their email addresses are 21d070090@iitb.ac.in, 22d0549@iitb.ac.in, and gskasbekar@ee.iitb.ac.in. }}

\begin{document}

\maketitle

\begin{abstract}
We design a lightweight authentication and key agreement protocol for Internet of Drones (IoD) environments that simultaneously provides post-quantum security, hardware-rooted trust, and resistance against machine-learning-based Physical Unclonable Function (PUF) modeling attacks. Existing PUF-based authentication schemes for unmanned aerial vehicles (UAVs) offer protection against cloning and physical capture, but generally lack resistance against quantum adversaries. Conversely, recent post-quantum UAV authentication protocols provide quantum-resistant key establishment, yet do not incorporate hardware-bound authentication mechanisms and therefore remain vulnerable to device cloning and physical compromise. To address these limitations, we integrate CRYSTALS-Kyber, dynamic challenge scrambling (DCS), fuzzy extractors, and PUF-based identity binding into a unified framework. The resulting protocol enables anonymous authentication, forward secrecy, resistance to physical capture and replay attacks, and secure drone-to-drone communication, while maintaining suitability for resource-constrained UAV platforms. We provide a  formal security analysis of the proposed scheme under the real-or-random model,  a ProVerif-based formal verification, an informal analysis of its effectiveness against various attacks, and a performance evaluation of the proposed scheme through communication, computation, and storage cost analyses, which show its advantages over state-of-the-art architectures.
\end{abstract}


\section{Introduction}
{The rapid advancement of the Internet of Drones (IoD) has transformed unmanned aerial vehicles (UAVs) from isolated aerial platforms into highly interconnected intelligent systems capable of autonomous coordination, surveillance, delivery, disaster management, and military operations \cite{mozaffari2019tutorial, shakhatreh2019unmanned}. Modern UAV ecosystems rely heavily on continuous communication between drones and ground stations (GS). This increasing interconnectivity has significantly expanded the attack surface of drone networks, making secure authentication and key establishment critical requirements for maintaining operational integrity and reliability. In particular, authentication failures in UAV environments can lead to catastrophic consequences, including unauthorized drone control, mission hijacking, data leakage, malicious swarm infiltration, and denial-of-service (DoS) attacks \cite{mall2022puf}.}

{Traditional authentication and key agreement protocols designed for conventional wireless networks are not well suited for UAV systems because drone networks operate under highly constrained conditions involving limited computational power, restricted battery capacity, unstable communication links, and dynamic topology changes. Furthermore, IoD systems are frequently deployed in hostile or unattended environments where drones are vulnerable to physical capture and invasive hardware attacks. To address this hardware-level threat, Physical Unclonable Functions (PUFs) have emerged as promising hardware-rooted trust primitives. A PUF exploits uncontrollable manufacturing variations in integrated circuits to generate unique device-specific responses that are practically impossible to duplicate \cite{pappu2002physical,gassend2002silicon}. Because PUF outputs are tied directly to physical hardware characteristics, they provide strong resistance against cloning and key extraction attacks. Consequently, several UAV authentication protocols have incorporated PUFs to achieve lightweight device authentication without permanently storing secret keys in memory \cite{xie2025lightweight,lounis2022d2d,tanveer2024paf}.}

{However, this reliance on symmetric cryptography, static credentials, or pre-stored challenge--response pairs (CRPs) exposes existing lightweight authentication schemes to replay attacks, cloning attacks, insider threats, and device impersonation \cite{mall2022puf}. In particular, protocols relying on static PUF CRP databases become vulnerable to machine learning-based modeling attacks once sufficient CRPs are collected by an adversary \cite{ruhrmair2010modeling}. Strong PUF architectures can often be approximated by attackers who collect large numbers of CRPs during protocol execution \cite{ruhrmair2010modeling}. Once a predictive model is trained, the attacker can emulate the behavior of the legitimate PUF without possessing the actual hardware. Recent advances in deep learning techniques have further amplified the practicality of such attacks, making static CRP-based authentication increasingly insecure for long-term deployment in IoD environments.}

{To overcome these limitations, recent research has proposed Dynamic Challenge Scrambling (DCS) mechanisms that obfuscate PUF challenges before evaluation. Instead of transmitting the real challenge directly, DCS dynamically transforms the challenge using session-dependent masking and permutation operations derived from ephemeral cryptographic material \cite{ebrahimabadi2024robust}. As a result, the challenge observed on the communication channel is never the actual challenge evaluated by the PUF. This significantly increases the difficulty of collecting meaningful CRPs for supervised machine learning attacks, since the observed challenge space becomes session-randomized and unpredictable.}

{At the same time, the emergence of large-scale quantum computing poses a serious threat to currently deployed public-key cryptographic systems. Widely used asymmetric algorithms such as Rivest–Shamir–Adleman (RSA) and elliptic curve cryptography (ECC) can be broken efficiently using Shor’s algorithm \cite{shor1994algorithms}, rendering many existing UAV authentication protocols insecure in the post-quantum era. Since UAV infrastructures are increasingly being integrated into long-term civilian, industrial, and defense applications, it is essential to design authentication frameworks that remain secure even against future quantum adversaries. This has motivated significant research interest in post-quantum cryptography (PQC), particularly lattice-based constructions such as the CRYSTALS-Kyber Key Encapsulation Mechanism (KEM), which was standardized by the National Institute of Standards and Technology for post-quantum secure key exchange \cite{alagic2024module}.}

{Existing UAV authentication protocols have addressed individual security challenges such as lightweight authentication, PUF-based device verification, anonymous communication, and post-quantum key exchange, but they continue to exhibit important limitations. Earlier PUF-based schemes for IoD environments \cite{xie2025lightweight,lounis2022d2d,tanveer2024paf,choi2025puf,ahmad2026secure,alkatheiri2022lightweight}  provide lightweight authentication and resistance against basic impersonation attacks, yet most rely on static challenge–response structures that remain vulnerable to machine-learning-based PUF modeling attacks and do not offer post-quantum security guarantees. Conversely, recent post-quantum UAV authentication frameworks based on lattice cryptography and Kyber or Dilithium primitives \cite{xia2024quantum,mishra2023quantum,aissaoui2023authenticating,xie2025puf,tawfeeq2025lightweight,he2025post,kim2026pqrid} achieve resistance against quantum adversaries, but they generally lack hardware-rooted trust and therefore remain susceptible to physical capture and device cloning attacks. Other lightweight IoD authentication schemes \cite{jan2021design,wang2023lightweight,garcia2023tesla,cho2020sentinel,yu2023laka,bhattarai2024IoD,pan2024privacy} either depend heavily on centralized GS involvement, incur excessive communication overhead, or fail to provide anonymity and forward secrecy simultaneously. No existing scheme, to our knowledge, simultaneously combines hardware-rooted PUF trust, resistance to PUF modeling attacks, and post-quantum key establishment within a single lightweight IoD authentication framework.}

{This paper proposes a lightweight, post-quantum secure, and PUF-enhanced authentication and key agreement protocol for IoD environments that closes this gap. The proposed framework integrates CRYSTALS-Kyber KEM, DCS, fuzzy extractors, pseudo-identities, and hardware-bound PUF authentication into a unified security architecture. Unlike conventional PUF-based schemes that store CRP databases or static secrets, the proposed protocol stores only commitment values and helper data, thereby minimizing the impact of physical device compromise. The integration of DCS further ensures that PUF challenges remain session-specific and resistant to modeling attacks. In addition, the use of Kyber-based ephemeral key establishment guarantees forward secrecy and long-term resistance against quantum-enabled adversaries.}

{The proposed scheme supports both drone–to-GS authentication and decentralized drone-to-drone (D2D) mutual authentication without requiring continuous GS involvement. To preserve privacy and prevent traceability, drones employ dynamically refreshed pseudo-identities (PIDs) derived from monotonic counters and cryptographic hashing. The protocol also incorporates a secure revocation mechanism using authenticated broadcasts and post-quantum signatures to permanently exclude compromised drones from the network. Formal security analysis under the real-or-random (RoR) model \cite{bellare1993entity} demonstrates that the proposed protocol achieves semantic session-key security under the hardness assumptions of module learning with errors (MLWE), secure hash functions, and PUF unpredictability. Furthermore, informal analysis shows resistance against replay attacks, impersonation attacks, physical capture attacks, desynchronization attacks, insider attacks, and man-in-the-middle attacks.}

{In contrast to prior approaches, the proposed protocol simultaneously achieves post-quantum security, resistance against PUF modeling attacks, hardware-rooted unclonability, anonymous authentication, forward secrecy, and decentralized drone-to-drone mutual authentication without continuous GS participation, by integrating CRYSTALS-Kyber KEM, PUF-based hardware authentication, DCS, fuzzy extractors, and dynamically refreshed pseudo-identities into a unified lightweight framework. To the best of our knowledge, very few existing IoD authentication protocols combine lattice-based post-quantum key establishment with dynamically obfuscated PUF authentication in a unified resource-efficient framework. Therefore, the proposed scheme provides a practical and future-proof security solution for next-generation UAV communication systems operating in adversarial and quantum-capable environments.}

The remainder of this paper is organized as follows. Section \ref{sec:related_works} reviews related work. Section \ref{sec:system_model} presents the system model, problem formulation, adversarial model, background on the MLWE problem, and the cryptographic primitives used. Section \ref{sec:proposed_protocol} describes the proposed protocol in detail across six phases. Section \ref{sec:formal_security} provides formal security analysis under the RoR model, followed by a ProVerif-based verification in Section \ref{sec:proverif}. Section \ref{sec:informal_sec_proof} provides an informal analysis of the effectiveness of the protocol against various  attacks. Section \ref{sec:performance_compare} evaluates the performance of the proposed scheme through communication, computation, and storage cost analyses, and compares it against state-of-the-art schemes. Section \ref{sec:conclusion} concludes the paper and outlines directions for future research.

\section{Related Work}\label{sec:related_works}
{The security of UAVs and IoD has attracted significant attention in recent years, with various schemes being proposed to counter impersonation, replay, and physical capture threats. Earlier lightweight designs primarily relied on PUFs for authentication. The work \cite{xie2025lightweight} introduced a PUF-enabled lightweight authentication protocol for drone--GS and drone--drone communication. The D2D-MAP protocol \cite{lounis2022d2d} uses PUF-based mutual authentication between drones, reducing reliance on any centralized entity. The work \cite{tanveer2024paf} proposed PAF-IoD, a PUF-enabled authentication framework for IoD systems that ties authentication secrets to PUFs at both the GS and drones, raising the difficulty of GS key extraction and reuse, and improving on earlier lightweight PUF schemes such as \cite{xie2025lightweight} that did not defend well against GS physical capture. More recently, \cite{xie2025puf} proposed a lightweight PUF-based anonymous cross-domain authentication protocol in which neither the GS nor the control center stores drone credentials, effectively eliminating stolen-verifier and privacy-leakage threats. In \cite{choi2025puf}, security vulnerabilities in a prior PUF-based IoD authentication scheme were identified and corrected, and an improved protocol with stronger resistance to physical capture and replay attacks was presented within a challenge-response framework. In \cite{ahmad2026secure}, a fully decentralized swarm authentication scheme was proposed, relying entirely on PUFs and true random number generators, with no pre-distributed keys and no central authority, resisting cloning and replay attacks through token evolution and statistical trust evaluation. All of these schemes \cite{xie2025lightweight,lounis2022d2d,tanveer2024paf,xie2025puf,choi2025puf,ahmad2026secure}, however, omit any post-quantum secure mechanism and thus remain vulnerable to quantum attacks \cite{shor1994algorithms}; moreover, those relying on static CRP structures remain susceptible to machine-learning-based PUF modeling attacks \cite{ruhrmair2010modeling}, and none of them offer strong forward secrecy guarantees. In contrast, our scheme maintains the benefits of PUF-based hardware trust and GS-free authentication while incorporating DCS to strengthen resistance against modeling attacks and post-quantum Kyber encapsulation to ensure forward secrecy.}

{The use of post-quantum cryptography in UAV networks has been investigated in parallel, generally without hardware-rooted trust anchors. In \cite{xia2024quantum}, lattice-based Kyber was applied to UAV communications, demonstrating the feasibility of quantum-resistant key exchange in such networks. In \cite{mishra2023quantum},  a quantum-safe protocol for IoD communications using lattice assumptions was designed. In \cite{aissaoui2023authenticating},  the use of Falcon and Dilithium post-quantum signatures for securing civil and military UAV communications was explored. While offering strong authentication, signature-based schemes suffer from high communication and storage overhead because of large signature sizes, limiting its suitability for more resource-constrained drones. Benchmark studies confirm the practicality of our design choices. In \cite{tawfeeq2025lightweight}, post-quantum KEMs (Kyber, BIKE, HQC) for UAV communication have been evaluated, concluding that Kyber offers the best performance trade-offs for constrained devices, a finding that supports our own adoption of Kyber as the primary key establishment mechanism. In \cite{he2025post}, a post-quantum authentication and key agreement scheme for drone swarms was proposed, combining Kyber-based group key agreement with a sparse Merkle tree (SMT) identity system supporting dynamic membership changes, with experimental results showing low resource overhead on embedded hardware. In \cite{kim2026pqrid}, pqRID, the first Remote ID-compliant, post-quantum authenticated identification system for drones was introduced, using pseudonym certificates and decoupled PQC-based authentication to satisfy Federal Aviation Administration (FAA) remote ID requirements while preserving flight-path privacy, with a reported 9\,ms authentication latency on a Raspberry Pi-mounted UAV. However, pqRID specifically targets broadcast identification rather than a full authenticated key agreement setting and does not address drone-to-drone authentication. Overall, these schemes \cite{xia2024quantum,mishra2023quantum,aissaoui2023authenticating,tawfeeq2025lightweight,he2025post,kim2026pqrid} do not integrate hardware trust anchors such as PUFs, leaving them vulnerable to device-cloning and physical capture attacks \cite{mall2022puf}; our approach improves upon these by combining lattice-based KEMs with device-bound PUF authentication, avoiding the overhead of post-quantum signatures in the authentication phase while still achieving hardware-rooted unclonability.}

{A further set of lightweight drone authentication schemes rely on classical cryptographic primitives, with some depending on blockchain infrastructure for decentralized trust management. In \cite{hussain2023user}, an efficient user access protocol was introduced, which reinforces overall system reliability in dynamic drone environments, providing mutual authentication for user and drone through one-way hash functions. In \cite{garcia2023tesla}, a $\mu$Tesla-based approach was combined with blockchain to protect broadcast communications within drone networks, but it relies on GS for authentication. SENTINEL \cite{cho2020sentinel} is a secure and efficient authentication framework emphasizing rapid and reliable authentication for dynamic drone networks. LAKA-UAV \cite{yu2023laka} is a lightweight authentication and key agreement scheme that leverages blockchain for cloud-assisted drone networks in flying ad-hoc environments. Their framework focuses on providing robust security with minimal computational overhead, making it suitable for dynamic and distributed drone communications. In \cite{bhattarai2024IoD},  a lightweight and anonymous, application-aware authentication and key agreement protocol was proposed, emphasizing user privacy and efficient credential management. In \cite{pan2024privacy}, a privacy-preserving, blockchain-based location authentication scheme for low-altitude UAVs, which guarantees auditability, was proposed; however, it incurs considerable computational overhead. All of these protocols \cite{hussain2023user,garcia2023tesla,cho2020sentinel,yu2023laka,bhattarai2024IoD,pan2024privacy} were found to be ineffective against physical capture attacks and did not provide the feature of anonymity. Also, \cite{garcia2023tesla,yu2023laka,pan2024privacy} rely on blockchain infrastructure and none offer post-quantum security or PUF-based hardware trust. The PUF-based authentication layer in our scheme can complement such blockchain-based frameworks by providing a lightweight, hardware-rooted trusted identity input for secure verification.}

{A separate line of work strengthens PUF hardware itself against machine-learning-based modeling attacks rather than proposing full authentication protocols. In \cite{ibrahim2024resilience},  memristor-based PUFs were evaluated under machine learning attacks, demonstrating that advanced PUF structures can withstand modeling better than SRAM counterparts. The LPUF-AuthNet framework \cite{mefgouda2024lpuf} incorporates machine-learning-aware defenses \cite{nguyen2018interpose} to make PUFs resistant to modeling while maintaining low inference cost. While such hardware-level improvements are valuable, neither prevents CRP exposure at the protocol level once challenges are observed on the communication channel. Our scheme instead strengthens the protocol layer: DCS ensures that even an easily modeled PUF cannot be exploited in operation, since challenges are scrambled session-by-session using ephemeral Kyber-derived keys, preventing the formation of exploitable CRP datasets regardless of the underlying PUF's hardware resilience.}

{Beyond UAV-specific protocols, \cite{han2024enhanced} proposed an enhanced multifactor authentication and key agreement protocol; however, it is not tailored to UAV networks and does not integrate post-quantum cryptography or PUF-based identity proofs. In \cite{khan2024future}, the challenges of future-proofing UAV communications with post-quantum cryptography were surveyed, emphasizing the need for lightweight, quantum-resistant schemes suitable for real-time UAV applications. Our scheme adapts these general ideas to the UAV setting, directly addressing this gap by integrating dynamic PUF obfuscation with post-quantum security to provide both PQC security and hardware-rooted unclonability.}

{Timestamp-free anonymous authentication models \cite{gupta2022secure} are useful in Internet of Things (IoT) systems lacking synchronized clocks. We adopt their principle by using counters and nonces instead of timestamps, thereby avoiding synchronization attacks while preserving message freshness.}

{In summary, existing schemes individually address post-quantum resistance, PUF-based hardware trust, anonymity, resistance to PUF modeling attacks, or drone-to-drone authentication, but no single protocol combines all of these properties. In contrast to all of the above, the proposed scheme unifies post-quantum Kyber key establishment, DCS-obfuscated PUF authentication, forward secrecy, anonymity, and drone-to-drone key agreement in a single lightweight protocol, closing all the identified gaps simultaneously.}

\section{System Model, Problem Formulation, and Background}\label{sec:system_model}
This section gives a detailed description of the system model, network architecture, problem formulation, design objectives, threat model, and background.

\subsection{Entities Involved} \label{sec:entities_involved}
The system contains three types of entities: GS, drones, and adversary. The GS is a resource-rich infrastructure node acting as a registry, directory, and revocation authority. It stores device commitments and metadata and performs enrollment. Each drone $D_i$, $i \in \{1, \dots,n\}$, is equipped with a PUF, and cryptographic support for Kyber KEM and Authenticated Encryption with Associated Data (AEAD) \cite{cremers2023automated}. It also maintains a monotonic counter $CC_i$ and minimal helper data in secure storage. The adversary, denoted $A$, follows the Dolev–Yao model \cite{dolev1983security}, controlling the wireless medium with abilities to eavesdrop, replay, inject, and modify messages, and can physically capture drones to attempt side-channel or invasive extraction.

\subsection{Network Model} \label{sec:network_model}
\begin{figure}[h!]
\centering
\begin{tikzpicture}[>=Stealth, scale=0.75, every node/.style={transform shape}]

\draw[thick, purple!30, dashed] (0,0) circle (4cm);

\node[draw, rectangle, thick, minimum width=2.5cm, minimum height=1.2cm, fill=gray!10] (gs) at (0,-2) {GS};

\node[circle, draw=black, fill=gray!10, minimum size=1.2cm] (d1) at (-3,0.5) {Drone};
\node[circle, draw=black, fill=gray!10, minimum size=1.2cm] (d2) at (0,2.5) {Drone};
\node[circle, draw=black, fill=gray!10, minimum size=1.2cm] (d3) at (2.5,1) {Drone};
\node[circle, draw=black, fill=gray!10, minimum size=1.2cm] (d4) at (5,2) {Drone};
\node[circle, draw=black, fill=gray!10, minimum size=1.2cm] (d5) at (5,-2) {Drone};

\draw[<->, thick, blue, dashed] (gs) -- (d1);
\draw[<->, thick, blue, dashed] (gs) -- (d2);
\draw[<->, thick, blue, dashed] (gs) -- (d3);

\draw[<->, thick, orange, dashed] (d2) -- (d3);
\draw[<->, thick, orange, dashed] (d4) -- (d5);

\begin{scope}[xshift=0cm, yshift=-5.5cm]
  \draw[<->, thick, orange, dashed] (-3,0) -- ++(1.5,0) node[midway, above, black] {\scriptsize Drone and drone AKA};
  \draw[<->, thick, blue, dashed] (3,0) -- ++(1.5,0) node[midway, above, black] {\scriptsize Drone and GS AKA};
\end{scope}

\end{tikzpicture}
\caption{Network model showing drone-to-GS and D2D communication. Dashed blue lines represent drone-to-GS AKA. Dashed orange lines represent D2D AKA.}
\label{fig:network_model}
\end{figure}
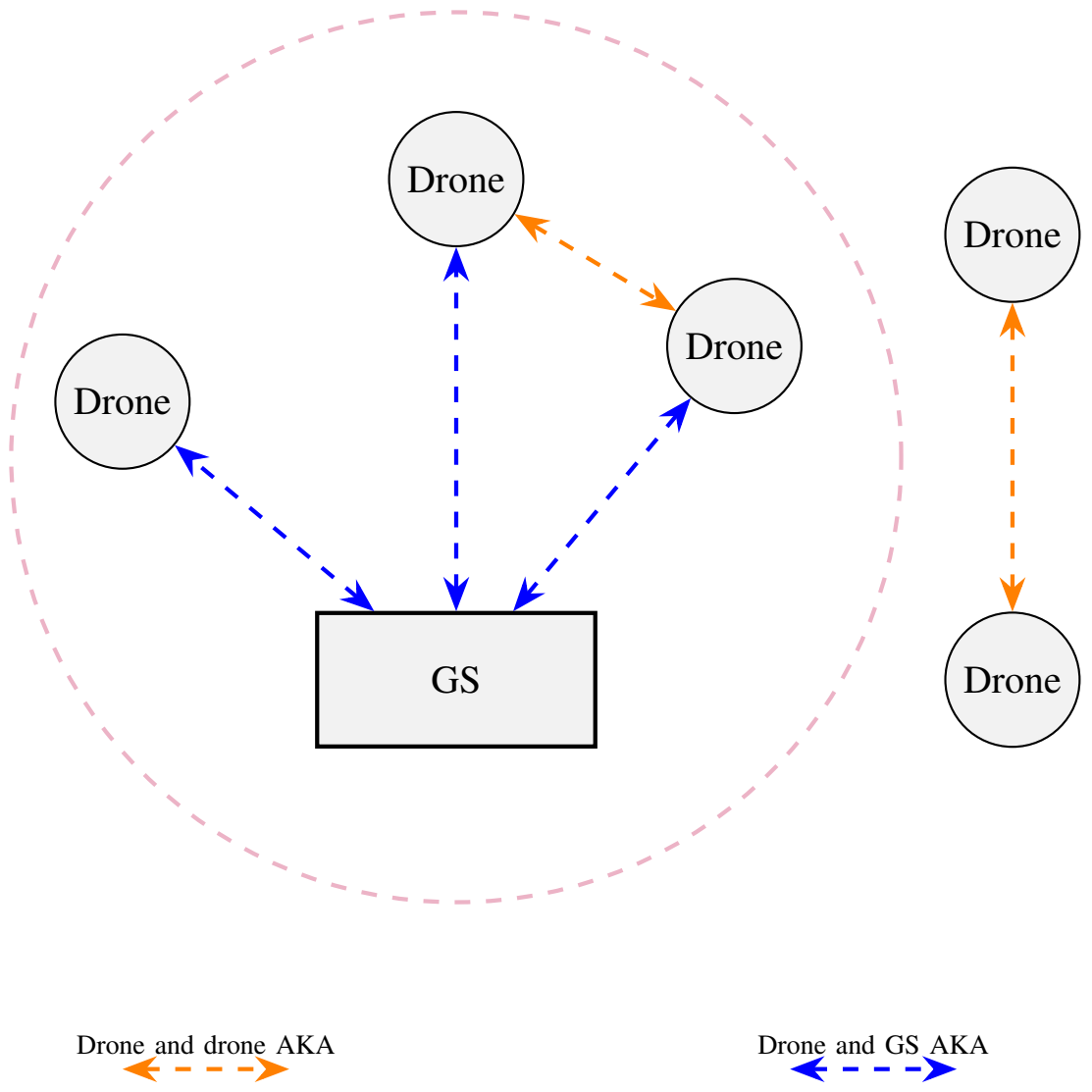

The network architecture consists of a GS and a fleet of drones $D_1, D_2, \dots, D_n$. These drones may communicate directly with each other or through the GS as necessary. The communication medium is public wireless, and could be unreliable and prone to packet loss and latency. When two drones do not know each other’s public keys, one drone may request the other’s ephemeral public key from the GS over an authenticated session, and the GS responds with a signed or AEAD-protected ephemeral key that can be cached temporarily to reduce communication overhead. Drones have limited storage, while the GS, though having abundant space, is restricted from storing raw PUF CRPs. Instead, it stores compact commitments and public keys so that even if compromised, the GS cannot be used to impersonate drones. Fig. \ref{fig:network_model} illustrates the network architecture. The dashed blue lines represent drone–GS authentication and key agreement (AKA) sessions, while the dashed orange lines represent D2D AKA sessions. A drone may communicate either directly with the GS or peer-to-peer with another drone, as shown in Fig. \ref{fig:network_model}. Table \ref{tab:notation} presents the notation used in the paper.

In this network model, we consider that drone $i$'s PUF is queried once with an enrollment challenge $Ch_i$ assigned by the GS. A fuzzy extractor \cite{dodis2004fuzzy} processes the raw PUF response to produce a stable secret $s_i$ and helper data $w_i$. The stable secret $s_i$ is used to compute a one-way commitment, $Com_i$, which is sent to and stored by the GS. Critically, $s_i$ is then erased from the drone. During flight operations, when the drone needs to authenticate to the GS, it generates a fresh Kyber key pair for that session. The drone then sends the ephemeral public key to the GS, and receives back a Kyber ciphertext. Only the drone possessing the matching ephemeral secret key can decapsulate this ciphertext to recover the session key. The GS verifies the drone's identity by checking whether the proof submitted by the drone is consistent with the stored commitment. This ensures that the GS holds only a hash, while the actual authentication secret lives only in the drone's silicon and is re-derived fresh each session.

\subsection{Objectives}
We seek to design an AKA protocol that supports lightweight operation, so that all cryptographic and PUF-related computations remain feasible on constrained UAV microcontrollers. The architecture should establish a hardware root of trust by binding each UAV’s identity to its silicon through a PUF, making cloning or key extraction prohibitively expensive. Post-quantum session keys should be ensured using Kyber KEM, providing quantum-resistant session establishment. The scheme should achieve anonymity by avoiding long-lived static identifiers and any dependence on synchronized timestamps. It should also ensures robustness against device capture by minimizing sensitive data storage, such as avoiding raw CRP tables at the GS, and by supporting straightforward revocation and reissue workflows. 

\subsection{Synchronization and Freshness Model} \label{sec:syn_freshness_model}
The system avoids reliance on global time synchronization across devices, as timestamp-based systems are susceptible to clock drift and synchronization attacks by adversaries. It ensures freshness through nonces and device-specific monotonic counters. Every communication session incorporates fresh nonces ($N, N_i, N_{GS}$) that are generated by the message originator and included in the AEAD associated data to guarantee uniqueness. Each drone $D_i$ maintains a persistent monotonic counter $CC_i$ that increments with every successful authentication. This counter helps generate PIDs that mask the true drone identity and detect replay or desynchronization attempts. Both the GS and drones maintain a small replay cache containing the identifiers of recent sessions (such as hashed tuples of $PID$, nonce, and KEM ciphertexts) to detect duplicates. This cache remains bounded to conform to UAV memory limitations while still preventing immediate replay.

\subsection{Public and Secret Key Management} \label{sec:public_secret_key_mgmt}
The key management system defines how cryptographic material is generated and managed across the network. {The GS possesses a long-term ML-DSA-44 key pair $(pk_{GS}, sk_{GS})$ \cite{national2024module} used for signing directory data and revocation broadcasts.} Each drone may hold a long-term key pair $(pk_i, sk_i)$, where $sk_i$ should be sealed using the onboard PUF or within a secure element (see Section \ref{SSC:PUF:Based:Key:Sealing}), ensuring that even if memory is extracted, impersonation remains infeasible. {In each session, drone $D_i$ sends a fresh ephemeral Kyber public key $pk_i^{ep}$ to its peer (the GS or drone $D_j$). The peer runs Kyber encapsulation under $pk_i^{ep}$ to obtain $(K_{sess}, c)$ and returns only the ciphertext $c$, denoted $c_{GS}$ or $c_j$. $D_i$ recovers $K_{sess}$ by decapsulating $c$ with $sk_i^{ep}$. The long-term key pairs are not used for encapsulation.} A group key $K_{grp}$ is also maintained for authenticating revocation broadcasts, periodically refreshed and redistributed under secure GS-drone sessions. The GS registry stores for each enrolled drone the set $(ID_i, pk_i, Com_i)$, where $Com_i$ is a one-way cryptographic commitment derived from the drone’s PUF.

\subsection{Physical Unclonable Function (PUF)} \label{sec:PUF}
Each drone’s PUF acts as a hardware root-of-trust, modeled as a function $PUF_i(Ch) \to Res$, mapping a challenge bitstring $Ch$ to a response $Res$ \cite{pappu2002physical, gassend2002silicon}. In our proposed scheme, we will use the Linear Feedback Shift Register (LFSR) PUF model. To address noise, a Fuzzy Extractor (FE)  \cite{dodis2004fuzzy} generates a stable secret $s_i$ and helper data $w_i$ from the raw response $(s_i, w_i) \leftarrow FE.Gen(Res)$. During operation, the drone can reproduce the same stable secret $s_i' = FE.Rep(Res', w_i)$ even from slightly noisy PUF responses. Only $w_i$ is stored locally, and neither the drone nor the GS retains $s_i$ in plaintext. {In authenticated sessions, the enrollment challenge $Ch_i$ is scrambled using an ephemeral KEM derived mask $K_{mask}$ before being input to the PUF, and the stable secret $s_i$ is subsequently reproduced from the resulting obfuscated response via the fuzzy extractor, ensuring that the PUF is always evaluated on a session specific, scrambled challenge and that the true challenge is never exposed on the communication channel.} This provides inherent resistance against modeling and cloning attacks.

A PUF serves as a hardware-based identity anchor that replaces the need to store secret keys in memory. The core insight is that all silicon chips, even those manufactured identically, develop microscopic random variations during fabrication, leading to differences in transistor switching thresholds, wire delays, and oxide thickness \cite{suh2007physical}. These variations are impossible to control during manufacturing and equally impossible to replicate afterward, even by the original manufacturer. When a challenge bitstring is applied to the PUF circuit, these physical variations determine the output response in a way that is unique to that specific chip. Two chips running the same challenge will produce different responses. This means the PUF response is the device’s identity. It cannot be copied to another chip, cannot be extracted as a file, and cannot be transferred digitally.
This is fundamentally different from storing a secret key in flash memory. A key stored in memory can be read out by an attacker who gains physical access to the chip. A PUF response, by contrast, is not stored anywhere. It only exists the moment the circuit is powered and the challenge is applied. Extracting it invasively (by decapping the chip and probing circuits) destroys the analog structures that generate the response, making the response permanently unavailable after such an attempt. This property is called physical uncloneability, and it is what makes PUFs attractive for resource-constrained devices like UAVs that operate in environments where physical capture is a realistic threat.


\subsection{PUF-Based Key Sealing Mechanism}
\label{SSC:PUF:Based:Key:Sealing}
\noindent

To ensure that long-term private keys are never stored in plaintext on the drone, the protocol uses a PUF-based key \textit{sealing} mechanism. Key sealing is a technique in hardware security \cite{10.1145/2818186}, where a cryptographic key is encrypted (``sealed'') under a stable secret derived from the device’s PUF. The sealed form is stored in memory, while the key can only be recovered (``unsealed'') by re-evaluating the PUF on the same device.

During enrollment (see Section \ref{SSC:drone:enrollment:phase}), after drone $D_i$ generates its long-term key pair $(pk_i, sk_i)$, it also derives a stable PUF secret $s_i$ via:
\[
(s_i, w_i) \leftarrow FE.Gen(PUF_i(Ch_i)).
\]
The drone then computes the sealed representation of its private key:
\[
Seal_i = AEAD_{s_i}(sk_i; AAD = ID_i),
\]
and stores only $\{ Seal_i, w_i \}$ in non-volatile memory. The plaintext private key $sk_i$ is erased. Because $s_i$ is never stored and can only be reproduced through:
\[
s_i' = FE.Rep(PUF_i(Ch_i), w_i),
\]
where $Ch_i$ is the challenge script assigned to drone $D_i$, an attacker cannot recover the sealed key without possessing the genuine hardware.

During authentication, the drone unseals its private key by reconstructing $s_i'$ and computing:
\[
sk_i = AEAD^{-1}_{s_i'}(Seal_i; AAD = ID_i).
\]
Thus, even if the sealed blob or helper data are extracted from memory, the key cannot be recovered on another device.

\subsection{Kyber Key Encapsulation Mechanism}

Kyber is a post-quantum KEM standardized by National Institute of Standards and Technology (NIST) in 2022 for public-key encryption and key exchange in the presence of quantum-capable adversaries \cite{alagic2024module}. It is based on the hardness of the MLWE problem (see Section \ref{SSC:MLWE}). MLWE is a lattice-based problem involving noisy linear equations defined over polynomial rings, and no known classical or quantum algorithm can solve it in polynomial-time, making Kyber one of the most reliable choices for security in long-term and high-risk environments such as the IoD.

It functions as a KEM rather than a direct encryption system. In a KEM, one party publishes a public key while keeping its matching secret key private. A second party uses the public key to generate two outputs: a ciphertext and a shared secret. The ciphertext encapsulates a random seed, and the shared secret is derived from the same seed. The receiving party decapsulates the ciphertext using its secret key to recover an identical shared secret. Because the shared key never appears in the communication channel, and only the ciphertext is transmitted, the KEM approach provides inherent protection against key-recovery attacks.

Kyber consists of two main procedures: \textit{encapsulation} and \textit{decapsulation}. A drone generates an ephemeral public-key pair $(pk, sk)$ and sends $pk$ to the GS. The GS performs Kyber encapsulation to obtain a ciphertext $c$ and a shared secret $K_{sess}$. The ciphertext is transmitted back to the drone. Using its secret key, the drone decapsulates the ciphertext to reproduce the same $K_{sess}$. At this point, both parties share an identical 256-bit session key derived from MLWE hardness, without exposing the randomness or secret material.

Kyber offers us various advantages in the area of UAV security. Firstly, it provides post-quantum forward secrecy: even if an adversary stores communication today and obtains a quantum computer in the future, they cannot recover the session keys. Also, Kyber is computationally efficient and memory-efficient, making it suitable for embedded drone hardware with limited resources.

\subsection{Dynamic Challenge Scrambling (DCS)} \label{sec:dcs}
\noindent

DCS is a lightweight protocol-level countermeasure designed to prevent machine-learning based modelling attacks on PUFs by ensuring that every on-air PUF query is effectively one-time and obfuscated. The key idea is simple: rather than sending a raw challenge bitstring \(Ch\) to the device, the verifier uses ephemeral session material to transform or permute the challenge bits so that the actual PUF input differs each time in a way that an eavesdropper cannot predict or reproduce. This defeats an adversary who attempts to collect a large, consistent set of CRPs for supervised learning, because the observed challenges on the wire are never the true challenges the PUF evaluates and the obfuscation pattern is different every session. The DCS concept was proposed and experimental validated in a recent work by Ebrahimabadi et al. \cite{ebrahimabadi2024robust}.

DCS first derives a pseudorandom stream from the KEM-secured session material and a freshness nonce, uses a prefix of that stream as a bitwise XOR mask, and uses the remainder to seed a lightweight permutation/bit-swap of the masked challenge. The permutation may be implemented as a sequence of pairwise bit-swaps or small block shuffles. Both the XOR mask and the permutation seed are unpredictable to an eavesdropper who does not know \(K_{mask}\).

\paragraph{Verifier/ Prover Message Flow (Integration Sketch)} In a GS--drone authentication round, the DCS steps are integrated as follows:

\begin{enumerate}
  \item The parties derive an ephemeral \(K_{sess}\) via Kyber KEM.
  \item The party that will submit the PUF evaluation constructs the nonce \(N\) and computes \(Ch'=\text{DCS\_Scramble}(Ch, K_{mask}, N)\).
  \item The party evaluates \(Res = \text{PUF}(Ch')\), runs a fuzzy extractor reproduction \(s=\text{FE.Rep}(Res,w)\) to stabilize the response, and then produces an AEAD-protected PUF proof bound to \(K_{sess}\) (e.g., HMAC or AEAD of \(H(Res)\) plus nonces).
  \item The verifier recomputes \(Ch'\) (it has \(K_{mask}\) and \(N\)) and checks the AEAD/HMAC proof. Because \(K_{mask}\) is unknown to an eavesdropper, the challenge on the wire is useless for model training.
\end{enumerate}
This flow couples DCS to Kyber-derived session material, so breaking the obfuscation requires either breaking Kyber (quantum-hard) or acquiring the live PUF (hardware capture).

\subsection{Module Learning With Errors}
\label{SSC:MLWE}
The security of the Kyber KEM used in the proposed protocol is based on the hardness of the MLWE problem \cite{langlois2012modulelattices}.  

Let
\[
R_q = \mathbb{Z}_q[X]/(X^n + 1)
\]
be the quotient polynomial ring, where \(q\) is a prime modulus and \(n\) is a power of two. Let \(k\) denote the module rank. The MLWE distribution \(A_{s,\chi}\) is defined as follows:

\begin{enumerate}
    \item Sample a secret vector
    \[    \mathbf{s} \in R_q^k \]
    uniformly at random.
    
    \item Sample a matrix
    \[
    \mathbf{A} \in R_q^{k \times k}
    \]
    uniformly at random.
    
    \item Sample an error vector
    \[
    \mathbf{e} \leftarrow \chi^k
    \]
    from a centered binomial distribution \(\chi\).
    
    \item Compute
    \[
    \mathbf{t} = \mathbf{A}\mathbf{s} + \mathbf{e}.
    \]
\end{enumerate}

\textit{Decision MLWE Problem:}  
Given \((\mathbf{A}, \mathbf{t})\), distinguish between whether $t$ was sampled from \(A_{s,\chi}\) or uniformly at random from \(R_q^k\).

\textit{Search MLWE Problem:}  
Given \((\mathbf{A}, \mathbf{t})\) sampled from \(A_{s,\chi}\), recover the secret vector $s$.

No known classical or quantum algorithm solves MLWE in polynomial time \cite{langlois2012modulelattices,shor1994algorithms}. For the Kyber-512 parameter set used in this paper,
\[
n = 256, \qquad k = 2, \qquad q = 3329.
\]

The advantage of any probabilistic polynomial-time (PPT) adversary \(\mathcal{A}\) in solving Decision MLWE is assumed to be negligible:
\[
\mathrm{Adv}^{\mathrm{MLWE}}_{\mathcal{A}} \leq \epsilon,
\]
where \(\epsilon\) is a negligible function.

This hardness assumption forms the foundation of the security proof in Theorem \ref{theorem1_statement} (Section \ref{sec:formal_sec_theorem_proof}), specifically through the game transition in \(\mathcal{G}_4\).

\subsection{Adversary Model}

We adopt the Dolev--Yao adversary model~\cite{dolev1983security}, which is the standard model for cryptographic protocol analysis. The adversary \(\mathcal{A}\) is modeled as a PPT algorithm with full control over the wireless communication channel. Specifically, \(\mathcal{A}\) can perform the following actions:

\begin{itemize}
    \item \textit{Eavesdrop:} Intercept all messages transmitted over the network.
    
    \item \textit{Replay:} Retransmit previously captured messages at a later time.
    
    \item \textit{Inject:} Insert arbitrary messages into the communication channel.
    
    \item \textit{Modify:} Alter intercepted messages before forwarding them.
    
    \item \textit{Physically capture:} Seize a drone and attempt to extract secrets from its non-volatile memory through side-channel or invasive attacks.
    
    \item \textit{Model PUFs:} Collect CRPs during protocol execution and train a machine-learning model to emulate PUF behavior.
\end{itemize}

We assume that \(\mathcal{A}\) cannot solve computationally hard problems in polynomial time, specifically the MLWE problem underlying the Kyber KEM (see Section \ref{SSC:MLWE}), and cannot clone PUF hardware. The GS is assumed to be a trusted entity during the enrollment phase. However, we additionally consider the scenario in which the GS is physically compromised after deployment and evaluate whether an adversary can extract useful secrets from the compromised system.

The formal adversary model used in the security proofs is the RoR model\cite{bellare1993entity}, which is described in detail in Section \ref{sec:formal_security}.

\section{Proposed Protocol}\label{sec:proposed_protocol}

\indent

\begin{table*}[t]
\centering
\caption{The notations used in the paper.}
\label{tab:notation}
\begin{tabular}{|l|l|}
\hline
\textbf{Notation} & \textbf{Description} \\
\hline
GS & Ground Station \\
$D_i, D_j$ & Drone $i$, drone $j$ \\
$ID_{GS}, ID_i, ID_j$ & Long-term identities of GS and drones \\
$PID_i^{(t)}, PID_j^{(t)}$ & Pseudo-identities used in session $t$ \\
$PID_i'$ & Refreshed PID of drone $i$ \\
$pk_{GS}, sk_{GS}$ & GS public/private key pair \\
$pk_i, sk_i$ & Drone $i$ public/private key pair (optional; often sealed) \\
$K_{grp}$ & Group symmetric key for revocation broadcasts \\
$RND()$ & Secure random number generator primitive \\
PUF$_i$(Ch) & Physical Unclonable Function of drone $i$ evaluated on challenge Ch \\
$Ch_i, Ch_{GS}$ & Enrollment challenge for drone $i$ and GS challenge \\
$C_i, C'_i$ & Original and scrambled challenge for drone $i$ (DCS) \\
$Res_i, R_i, R'_i, Res_{GS}$ & Raw and scrambled PUF responses (drone $i$ and GS) \\
$FE.Gen, FE.Rep$ & Fuzzy extractor to generate/ reproduce functions \\
$s_i, w_i$ & Stable secret and helper data from FE.Gen(Res) \\
$Com_i$ & Commitment stored at GS for drone $i$ \\
$M_i$ & Random mask used in Dynamic Challenge Scrambling (DCS) \\
$K_{mask}$ & Mask key derived from session material for DCS \\
Kyber.Encap($pk$), Kyber.Decap($c$, $sk$) & Kyber-512 encapsulation and decapsulation operations \\
$c_{GS}, c_j$ & Kyber ciphertexts sent by the GS and drone $D_j$ \\
$K_{sess}, K_{temp}, K_{AB}$ & Derived session/temporary keys \\
$AEAD_K(M;AAD)$ & Authenticated encryption of message $M$ under key $K$ \\
$H(\cdot)$ & Cryptographic hash function (e.g., SHA-256) \\
$HMAC_K(\cdot)$ & Hash-based MAC with key $K$ \\
$N, N_i, N_{GS}, N_A, N_B$ & Fresh nonces generated by an entity \\
$CC_i$ & Monotonic counter on drone $i$ for PID updates \\
$Mask$ & HMAC-derived mask \\
$CT_A, CT_B$ & AEAD ciphertexts in D2D mutual authentication \\
$proof_A, proof_B$ & HMAC proofs binding PUF-derived values to session \\
$H\_res_A, H\_res_B$ & Hashed PUF responses used in proofs \\
$A_1, B_1$ & Public ephemeral values \\
$A_2, B_2, B_3$ & Authentication hash fields in AKA flows \\
$Q_i, Q_s$ & Obfuscated stored values combined with PUF responses \\
$epoch$ & Revocation epoch number \\
$\sigma_{GS}$ & {ML-DSA-44 signature of the GS on revocation messages} \\

\hline
\end{tabular}

\end{table*}

The proposed scheme is organized into six phases: (1) System initialization, (2) Drone
enrollment, (3) Drone-GS authentication and key agreement, (4) Drone-drone authentication and key agreement, (5) PID and public key update, (6) Drone revocation phase.

\subsection{System Initialization Phase}
\noindent

 The GS begins by generating its permanent identity $ID_{GS}$ and a corresponding long-term key pair $(pk_{GS}, sk_{GS})$. It then selects a secure post-quantum KEM (Kyber) and defines the system’s cryptographic primitives: a collision-resistant hash function $H(\cdot)$, an HMAC-based Key Derivation Function - $HKDF(\cdot)$, and an authenticated encryption primitive $AEAD_K(\cdot)$.

The GS then internally evaluates its onboard PUF, $PUF_{GS}$. It generates a random binary challenge $Ch_{GS}$ and computes the response $Res_{GS} = PUF_{GS}(Ch_{GS})$. It then extracts a stable internal secret $s_{GS}$ from $Res_{GS}$ using a fuzzy extractor $(s_{GS}, w_{GS}) \leftarrow FE.Gen(Res_{GS})$. The helper data $w_{GS}$ is stored securely, while $s_{GS}$ acts as the GS’s intrinsic hardware-bound secret for later identity binding. Unlike earlier PUF-only designs that retained raw challenge–response tables, the GS discards $Res_{GS}$ after extracting $s_{GS}$ to minimize physical capture exposure.

The GS also initializes the DCS function that will be used by all drones in subsequent sessions. It defines a scrambling procedure $C' = C \oplus H(K_{mask} \| N_{GS})$, where $K_{mask}$ is derived from an ephemeral session key $K_{sess}$ during authentication.

{The GS also constructs the master revocation and update key $K_{grp}$, which will be used to encrypt broadcast control messages such as revocation lists and reissue notices. These messages are additionally signed with the GS long-term ML-DSA-44 key $sk_{GS}$.} It also initializes the global revocation list $R$ as empty and the system epoch counter to zero. The GS then prepares the public system parameter set:
\begin{equation}
\begin{aligned}
\mathcal{P} = \{ &ID_{GS},\ pk_{GS},\ H(\cdot),\ HKDF(\cdot), AEAD, \\
&\ \text{Kyber parameters},\text{DCS function} \}.
\end{aligned}
\label{EQ:system:parameter:set}
\end{equation}
These parameters are transmitted to every newly enrolled UAV via a secure provisioning channel during enrollment. It is assumed that a secure channel is available only during enrollment.

At the end of the initialization phase, the GS maintains only its long-term key pair $(pk_{GS}, sk_{GS})$, helper data $w_{GS}$, and the internal hardware secret $s_{GS}$ sealed within its trusted hardware. No plaintext CRPs or unmasked PUF outputs are retained.

\subsection{Drone Enrollment Phase}
\label{SSC:drone:enrollment:phase}
\noindent

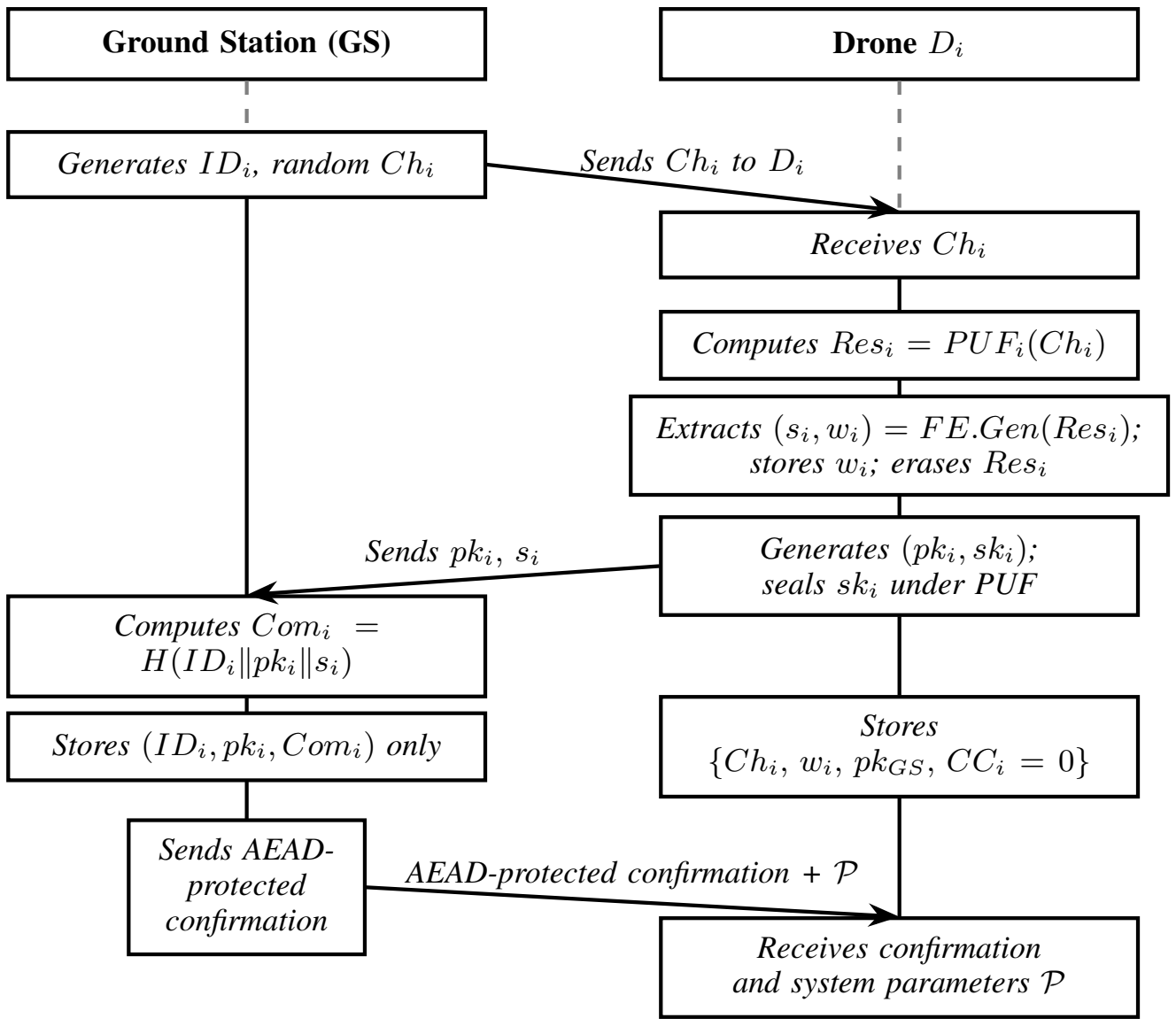
\begin{figure}[t]
\begin{tikzpicture}[
    scale=0.75,
    transform shape,
    >=Stealth,
    thick,
    stepbox/.style={
        rectangle, draw, thick,
        text width=4.4cm, align=center,
        minimum height=0.65cm,
        font=\small\itshape,
        inner sep=5pt
    },
    hdrbox/.style={
        rectangle, draw, thick,
        text width=4.4cm, align=center,
        minimum height=0.7cm,
        font=\small\bfseries,
        inner sep=5pt
    }
]

\def\GS{0}
\def\DR{6.5}

\node[hdrbox] (hGS) at (\GS, 0)  {\textbf{Ground Station (GS)}};
\node[hdrbox] (hDR) at (\DR, 0)  {\textbf{Drone $D_i$}};

\node[stepbox] (gsGen) at (\GS, -1.2) {Generates $ID_i$, random $Ch_i$};

\node[stepbox] (drRecv) at (\DR, -2) {Receives $Ch_i$};

\draw[->] (gsGen.east) -- node[above, font=\small\itshape]{Sends $Ch_i$ to $D_i$} (drRecv.north);

\draw[dashed, gray] (hGS.south) -- (gsGen.north);
\draw[dashed, gray] (hDR.south) -- (drRecv.north);

\node[stepbox] (drRes) at (\DR, -3) {Computes $Res_i = PUF_i(Ch_i)$};

\node[stepbox, text width=5cm] (drFE) at (\DR, -4)
    {Extracts $(s_i, w_i) = FE.Gen(Res_i)$; stores $w_i$; erases $Res_i$};

\node[stepbox] (drKey) at (\DR, -5.2)
    {Generates $(pk_i, sk_i)$; seals $sk_i$ under PUF};

\node[stepbox] (gsCom) at (\GS, -6)
    {Computes $Com_i = H(ID_i \| pk_i \| s_i)$};

\draw[->] (drKey.west) -- node[above, font=\small\itshape]{Sends $pk_i,\, s_i$} (gsCom.north);

\node[stepbox] (gsStore) at (\GS, -7)
    {Stores $(ID_i, pk_i, Com_i)$ only};

\node[stepbox] (drStore) at (\DR, -7)
    {Stores $\{Ch_i,\, w_i,\, pk_{GS},\, CC_i = 0\}$};

\node[stepbox, text width=2cm] (gsSend) at (\GS, -8.4)
    {Sends AEAD-protected confirmation};

\node[stepbox] (drConf) at (\DR, -9.2)
    {Receives confirmation and system parameters $\mathcal{P}$};

\draw[->] (gsSend.east) -- node[above, font=\small\itshape]{AEAD-protected confirmation + $\mathcal{P}$} (drConf.north);

\draw[-] (drRecv.south) -- (drRes.north);
\draw[-] (drRes.south)  -- (drFE.north);
\draw[-] (drFE.south)   -- (drKey.north);
\draw[-] (drKey.south)   -- (drStore.north);
\draw[-] (drStore.south)   -- (drConf.north);

\draw[-] (gsGen.south)  -- (gsCom.north);
\draw[-] (gsCom.south)  -- (gsStore.north);
\draw[-] (gsStore.south)  -- (gsSend.north);

\end{tikzpicture}
\caption{Drone $D_i$ Enrollment Phase.}
\label{fig:enrollment}
\end{figure}

The enrollment of each drone $D_i$ is performed once during provisioning and establishes the PUF-derived trust anchor, commitment, and long-term parameters required for future authentication. The following steps are used (see Fig. \ref{fig:enrollment}):

\begin{enumerate}
    \item GS assigns a unique long-term identity $ID_i$ to
    drone $D_i$ for internal registry purposes. It then selects a random enrollment challenge $Ch_i$ and transmits it securely to the drone.

    \item Drone $D_i$ inputs $Ch_i$ into its onboard PUF and computes the raw response:
    \[
        Res_i = PUF_i(Ch_i).
    \]

    \item Drone applies the fuzzy extractor to extract a stable hardware-bound secret:
    \[
        (s_i, w_i) \leftarrow FE.Gen(Res_i).
    \]
    The helper data $w_i$ is stored on the drone; the raw PUF response $Res_i$ is discarded.

    \item The Drone also generates a long-term key pair $(pk_i, sk_i)$ and seals $sk_i$ using its PUF or secure element.

    \item Drone erases $s_i$ from memory; it will later be reproduced using $FE.Rep$ when required. During authentication, `FE.Rep' uses the stored helper data $w_i$ to reconstruct the same stable value $s_i$ from a fresh (possibly noisy) PUF response, exploiting the error-correction property of the fuzzy extractor \cite{dodis2004fuzzy}. This ensures that $s_i$ is never stored in plaintext while remaining reproducible on demand from the live PUF hardware.

    \item Drone computes the commitment:
    \[
        Com_i = H(ID_i \,\|\, pk_i \,\|\, s_i),
    \]
    binding the drone identity and its PUF-derived secret and sends this commitment to the GS.

    \item GS stores only $(ID_i, pk_i, Com_i)$ in its registry.  
          No CRPs, no PUF responses, and no $s_i$ are ever stored.

    \item Drone stores $\{Ch_i, w_i, pk_{GS}, CC_i = 0\}$;  
          $sk_i$ is sealed; $s_i$ is never stored.

    \item Enrollment concludes with GS sending an AEAD-protected confirmation to $D_i$ containing system parameters $\mathcal{P}$  (see \eqref{EQ:system:parameter:set}) and acknowledgment of successful registration.
\end{enumerate}


\subsection{Drone--GS Authentication Phase}
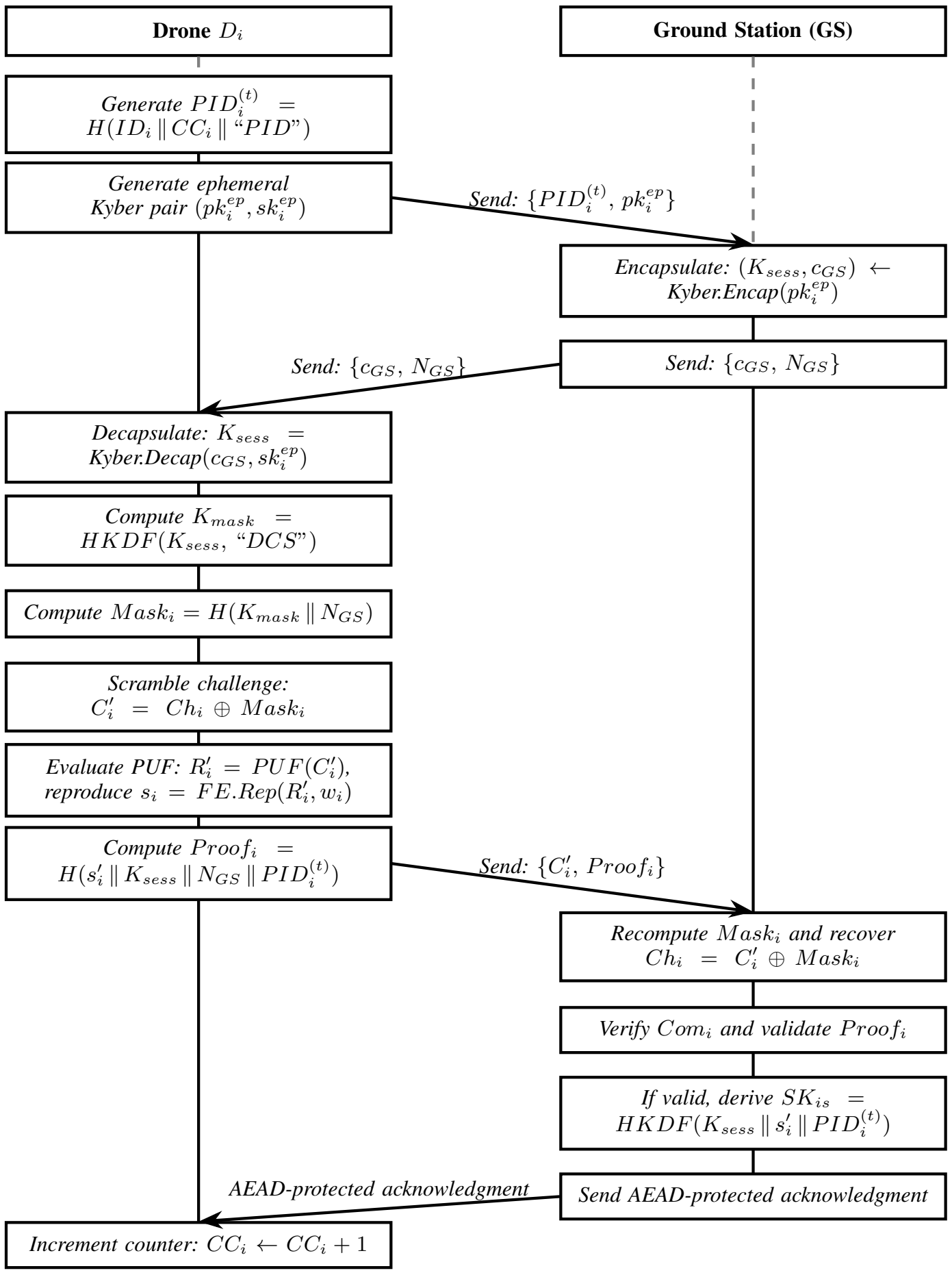
\begin{figure}[t]
\begin{tikzpicture}[
    scale=0.67,
    transform shape,
    >=Stealth,
    thick,
    stepbox/.style={
        rectangle, draw, thick,
        text width=5.2cm, align=center,
        minimum height=0.65cm,
        font=\small\itshape,
        inner sep=5pt
    },
    hdrbox/.style={
        rectangle, draw, thick,
        text width=5.2cm, align=center,
        minimum height=0.7cm,
        font=\small\bfseries,
        inner sep=5pt
    }
]

\def\DR{0}
\def\GS{8}

\node[hdrbox] (hDR) at (\DR, 0)  {\textbf{Drone $D_i$}};
\node[hdrbox] (hGS) at (\GS, 0)  {\textbf{Ground Station (GS)}};

\node[stepbox] (drPID) at (\DR, -1.2)
    {Generate $PID_i^{(t)} = H(ID_i \,\|\, CC_i \,\|\, ``PID")$};

\node[stepbox] (drEph) at (\DR, -2.4)
    {Generate ephemeral Kyber pair $(pk_i^{ep}, sk_i^{ep})$};

\node[stepbox] (gsSend1) at (\GS, -3.6) 
    {Encapsulate: $(K_{sess}, c_{GS}) \leftarrow \text{Kyber.Encap}(pk_i^{ep})$};
\draw[->] (drEph.east) -- node[above, font=\small\itshape]{Send: $\{ PID_i^{(t)},\, pk_i^{ep} \}$} (gsSend1.north);

\node[stepbox] (gsSend2) at (\GS, -4.8)
    {Send: $\{ c_{GS},\, N_{GS} \}$};

\node[stepbox] (drDecap) at (\DR, -6.0)
    {Decapsulate: $K_{sess} = \text{Kyber.Decap}(c_{GS}, sk_i^{ep})$};
\draw[->] (gsSend2.west) -- node[above, font=\small\itshape]{Send: $\{ c_{GS},\, N_{GS} \}$} (drDecap.north);

\node[stepbox] (drKmask) at (\DR, -7.2)
    {Compute $K_{mask} = HKDF(K_{sess},\, ``DCS")$};

\node[stepbox] (drMask) at (\DR, -8.4)
    {Compute $Mask_i = H(K_{mask} \,\|\, N_{GS})$};

\node[stepbox] (drScramble) at (\DR, -9.6)
    {Scramble challenge: $C'_i = Ch_i \oplus Mask_i$};

\node[stepbox] (drPUF) at (\DR, -10.8)
    {Evaluate PUF: $R'_i = PUF(C'_i)$, reproduce $s_i = FE.Rep(R'_i, w_i)$};

\node[stepbox] (drProof) at (\DR, -12.0)
    {Compute $Proof_i = H(s'_i \,\|\, K_{sess} \,\|\, N_{GS} \,\|\, PID_i^{(t)})$};

\node[stepbox] (gsRecomp) at (\GS, -13.2)
    {Recompute $Mask_i$ and recover $Ch_i = C'_i \oplus Mask_i$};
\draw[->] (drProof.east) -- node[above, font=\small\itshape]{Send: $\{ C'_i,\, Proof_i \}$} (gsRecomp.north);

\node[stepbox] (gsVerify) at (\GS, -14.4)
    {Verify $Com_i$ and validate $Proof_i$};

\node[stepbox] (gsDerive) at (\GS, -15.6)
    {If valid, derive $SK_{is} = HKDF(K_{sess} \,\|\, s'_i \,\|\, PID_i^{(t)})$};

\node[stepbox] (gsACK) at (\GS, -16.8)
    {Send AEAD-protected acknowledgment};

\node[stepbox] (drCC) at (\DR, -17.5)
    {Increment counter: $CC_i \leftarrow CC_i + 1$};
\draw[->] (gsACK.west) -- node[above, font=\small\itshape]{AEAD-protected acknowledgment} (drCC.north);

\draw[dashed, gray] (hDR.south)      -- (drPID.north);
\draw[dashed, gray] (hGS.south)      -- (gsSend1.north);

\draw[-] (drPID.south)      -- (drEph.north);
\draw[-] (drEph.south)      -- (drDecap.north);
\draw[-] (drDecap.south)    -- (drKmask.north);
\draw[-] (drKmask.south)    -- (drMask.north);
\draw[-] (drMask.south)     -- (drScramble.north);
\draw[-] (drScramble.south) -- (drPUF.north);
\draw[-] (drPUF.south)      -- (drProof.north);
\draw[-] (drProof.south)      -- (drCC.north);

\draw[-] (gsSend1.south)  -- (gsSend2.north);
\draw[-] (gsSend2.south)  -- (gsRecomp.north);
\draw[-] (gsRecomp.south) -- (gsVerify.north);
\draw[-] (gsVerify.south) -- (gsDerive.north);
\draw[-] (gsDerive.south) -- (gsACK.north);

\end{tikzpicture}
\caption{Drone--GS Authentication.}
\label{fig:drone_gs_auth}
\end{figure}

This phase  enables a drone $D_i$ to authenticate itself to the GS using a combination of Kyber post-quantum key exchange, DCS, PUF-based verification, and unlinkable PIDs. Both entities refresh their session secrets during every execution (see Fig. \ref{fig:drone_gs_auth}).

\begin{enumerate}
    \item Drone $D_i$ begins by deriving its current PID:
    \[
        PID_i^{(t)} = H(ID_i \,\|\, CC_i \,\|\, \text{``PID"}).
    \]
    It then generates a Kyber keypair $(pk_i^{ep}, sk_i^{ep})$ for this session and sends $PID_i^{(t)}$ together with $pk_i^{ep}$ to the GS.

    \item Upon receiving the ephemeral public key, the GS performs Kyber encapsulation:
    \[
       (K_{sess}, c_{GS}) \leftarrow \text{Kyber.Encap}(pk_i^{ep}),
    \]
    producing an ephemeral ciphertext $c_{GS}$ and shared secret $K_{sess}$. The GS sends $c_{GS}$ along with its own freshly generated nonce $N_{GS}$.

    \item Drone $D_i$ decapsulates the received ciphertext to recover the same shared secret using:
    \[
        K_{sess} = \text{Kyber.Decap}(c_{GS}, sk_i^{ep}).
    \]
    Both sides now share a post-quantum session key.

    \item Drone $D_i$ derives the DCS mask:
    \[
        K_{mask} = HKDF(K_{sess}, \text{``DCS"}),
    \]
    \[
        Mask_i = H(K_{mask} \,\|\, N_{GS}).
    \]
    It scrambles its stored enrollment challenge:
    \[
        C'_i = Ch_i \oplus Mask_i.
    \]
    Drone $D_i$ computes $R'_i = PUF_i(C'_i)$, reproduces $s_i = FE.Rep(R'_i, w_i)$, and generates the authentication proof:
    \[
        Proof_i = H(R'_i \,\|\, K_{sess} \,\|\, N_{GS} \,\|\, PID_i^{(t)}).
    \]
    The drone sends $\{C'_i, Proof_i\}$ to the GS.

    \item The GS computes the same mask $Mask_i$, recovers:
    \[
        Ch_i = C'_i \oplus Mask_i,
    \]
    and recomputes the expected PUF value using enrollment data:
    \[
        Expected = H(ID_i \,\|\, pk_i \,\|\, s_i),
    \]
    comparing it with the stored commitment $Com_i$. It then checks whether $Proof_i$ matches the reconstructed hash using its own recomputed $R'_i$.

    \item If the verification succeeds, the GS finalizes authentication by deriving the session key for secure communication:
    \[
        SK_{is} = HKDF(K_{sess} \,\|\, R'_i \,\|\, PID_i^{(t)}).
    \]
    The GS returns an AEAD-protected acknowledgment indicating successful authentication.  
    The drone increments its counter: $CC_i \leftarrow CC_i + 1$.
\end{enumerate}

\subsection{Drone--Drone Mutual Authentication and Key Exchange Phase}
\noindent

\begin{figure*}[t]
\centering
\begin{tikzpicture}[
    scale=0.75,
    transform shape,
    >=Stealth,
    thick,
    stepbox/.style={
        rectangle, draw, thick,
        text width=5.2cm, align=center,
        minimum height=0.65cm,
        font=\small\itshape,
        inner sep=5pt
    },
    hdrbox/.style={
        rectangle, draw, thick,
        text width=5.2cm, align=center,
        minimum height=0.7cm,
        font=\small\bfseries,
        inner sep=5pt
    }
]

\def\DI{0}   
\def\DJ{8}   

\node[hdrbox] (hDI) at (\DI, 0) {\textbf{Drone $D_i$ (Initiator)}};
\node[hdrbox] (hDJ) at (\DJ, 0) {\textbf{Drone $D_j$ (Responder)}};

\node[stepbox] (diPID) at (\DI, -1.2)
    {Generate $PID_i^{(t)} = H(ID_i \,\|\, CC_i \,\|\, ``PID")$};

\node[stepbox] (diEph) at (\DI, -2.4)
    {Generate ephemeral $(pk_i^{ep}, sk_i^{ep})$};

\node[stepbox] (djEncap) at (\DJ, -3.6)
    {Compute $(K_{sess}, c_j) \leftarrow Kyber.Encap(pk_i^{ep})$};
\draw[->] (diEph.east) -- node[above, font=\small\itshape]{Send $\{ PID_i^{(t)},\, pk_i^{ep} \}$} (djEncap.north);

\node[stepbox] (djNonce) at (\DJ, -4.8)
    {Choose $N_j$ and derive $K_{mask} = HKDF(K_{sess},\, ``DCS")$};

\node[stepbox] (djScramble) at (\DJ, -6.0)
    {Compute $C'_j = Ch_j \oplus H(K_{mask} \,\|\, N_j)$};

\node[stepbox] (djProof) at (\DJ, -7.2)
    {Compute $Proof_j = H(K_{sess} \,\|\, PID_i^{(t)} \,\|\, PID_j^{(t)} \,\|\, N_j)$};

\node[stepbox] (diDecap) at (\DI, -8.4)
    {Compute $K_{sess} = Kyber.Decap(c_j, sk_i^{ep})$};
\draw[->] (djProof.west) -- node[above, font=\small\itshape]{Send $\{ PID_j^{(t)},\, c_j,\, C'_j,\, N_j,\, Proof_j \}$} (diDecap.north);

\node[stepbox] (diKmask) at (\DI, -9.6)
    {Derive $K_{mask}$ and recover $Ch_j$};

\node[stepbox] (diPUF) at (\DI, -10.8)
    {Evaluate $Res_i' = PUF(Ch_j)$};

\node[stepbox] (diH) at (\DI, -12.0)
    {Compute $H_i = H(s_i')$};

\node[stepbox] (diProof) at (\DI, -13.2)
    {Compute $Proof_i = H(H_i \,\|\, K_{sess} \,\|\, PID_i^{(t)} \,\|\, N_j)$};

\node[stepbox] (djValidate) at (\DJ, -14.4)
    {Validate $Proof_i$};
\draw[->] (diProof.east) -- node[above, font=\small\itshape]{Send $\{ H_i,\, Proof_i \}$} (djValidate.north);

\node[stepbox] (djPUF) at (\DJ, -15.6)
    {Evaluate $Res_j' = PUF(Ch_i)$};

\node[stepbox] (djH) at (\DJ, -16.8)
    {Compute $H_j = H(s_j')$};

\node[stepbox] (djProofP) at (\DJ, -18.0)
    {Compute $Proof_j' = H(H_j \,\|\, K_{sess} \,\|\, PID_j^{(t)} \,\|\, N_i)$};

\node[stepbox] (diVerify) at (\DI, -19.2)
    {Verify $Proof_j'$};
\draw[->] (djProofP.west) -- node[above, font=\small\itshape]{Send $\{ H_j,\, Proof_j' \}$} (diVerify.north);

\node[stepbox, text width=6.5cm] (diDerive) at (\DI, -20.4)
    {Derive $K_{ij} = HKDF(K_{sess} \,\|\, H_i \,\|\, H_j \,\|\, PID_i^{(t)} \,\|\, PID_j^{(t)} \,\|\, ``FINAL")$};

\node[stepbox] (diCC) at (\DI, -21.6)
    {Update $CC_i$};

\node[stepbox] (djCC) at (\DJ, -21.6)
    {Update $CC_j$};

\draw[dashed, gray] (hDI.south) -- (diPID.north);
\draw[dashed, gray] (hDJ.south) -- (djEncap.north);

\draw[-] (diPID.south)    -- (diEph.north);
\draw[-] (diEph.south)    -- (diDecap.north);
\draw[-] (diDecap.south)  -- (diKmask.north);
\draw[-] (diKmask.south)  -- (diPUF.north);
\draw[-] (diPUF.south)    -- (diH.north);
\draw[-] (diH.south)      -- (diProof.north);
\draw[-] (diProof.south)  -- (diVerify.north);
\draw[-] (diVerify.south) -- (diDerive.north);
\draw[-] (diDerive.south) -- (diCC.north);

\draw[-] (djEncap.south)   -- (djNonce.north);
\draw[-] (djNonce.south)   -- (djScramble.north);
\draw[-] (djScramble.south)-- (djProof.north);
\draw[-] (djProof.south) -- (djValidate.north);
\draw[-] (djValidate.south)-- (djPUF.north);
\draw[-] (djPUF.south)     -- (djH.north);
\draw[-] (djH.south)       -- (djProofP.north);
\draw[-] (djProofP.south) -- (djCC.north);

\end{tikzpicture}
\caption{Drone--Drone Mutual Authentication and Key Exchange.}
\label{fig:drone_drone_new}
\end{figure*}
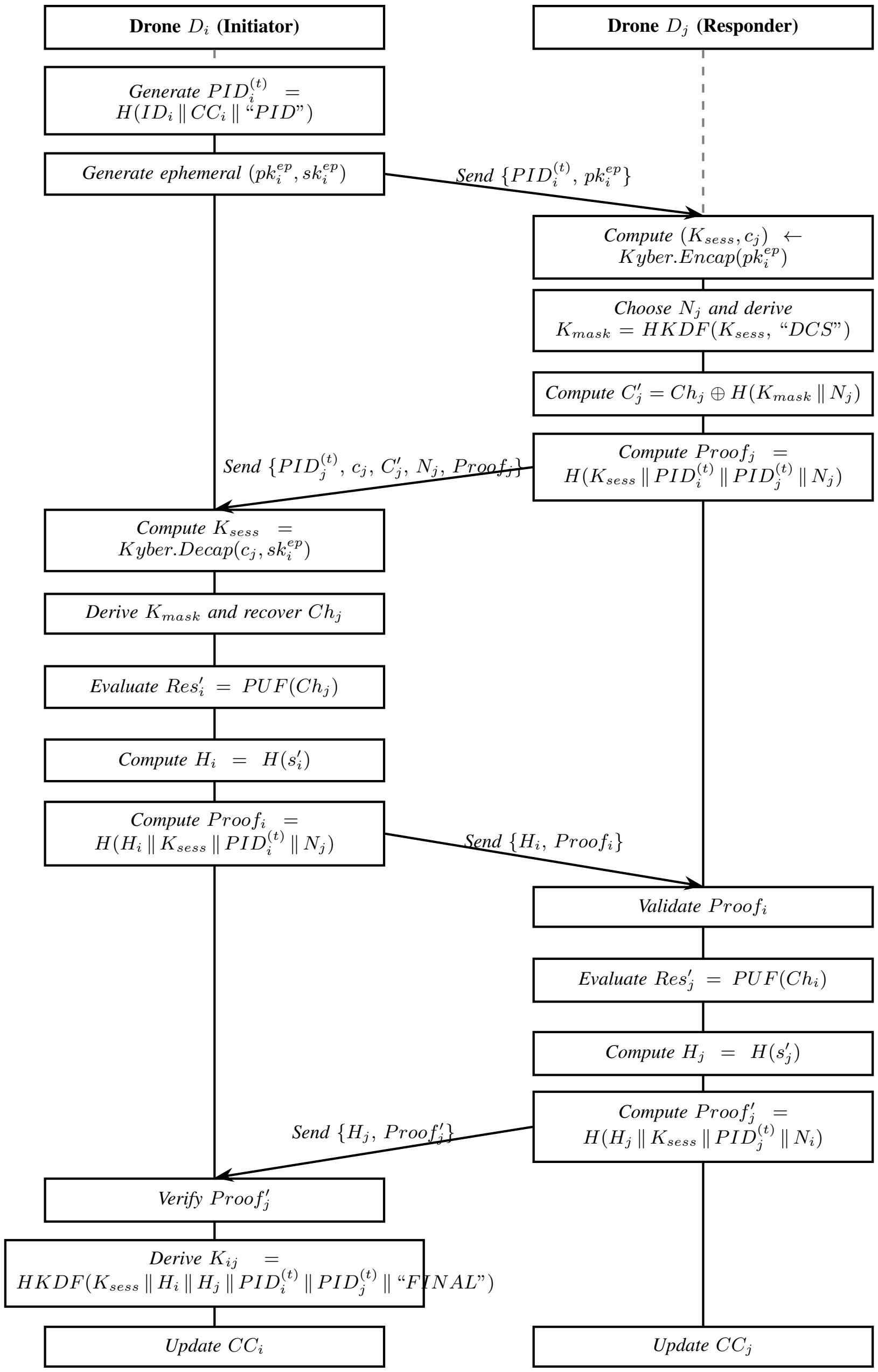

In this phase, two drones $D_i$ and $D_j$ mutually authenticate and derive a fresh post-quantum secure session key without relying on the GS. Each drone stores a long-term signing key sealed under its PUF, ensuring that extracting flash memory is insufficient for impersonation without access to the genuine silicon PUF. Uniqueness and freshness are enforced using nonces, PIDs and ephemeral keys. The detailed protocol is described below (see Fig. \ref{fig:drone_drone_new}).

\begin{enumerate}

\item \textbf{$D_i \rightarrow D_j$: Ephemeral KEM initiation}

Drone $D_i$ generates a new PID  
\[
PID_i^{(t)} = H(ID_i \,\|\, CC_i \,\|\, \text{``PID"}),
\]
then generates an ephemeral Kyber keypair $(pk_i^{ep}, sk_i^{ep})$ and sends  
\[
\{ PID_i^{(t)}, pk_i^{ep} \}
\]
to drone $D_j$.

\item \textbf{$D_j \rightarrow D_i$: Kyber encapsulation and scrambled challenge}

Drone $D_j$ computes the encapsulation  
\[
(K_{sess}, c_j) \leftarrow Kyber.Encap(pk_i^{ep}),
\]
selects a fresh nonce $N_j$, derives the DCS mask key  
\[
K_{mask} = HKDF(K_{sess}, \text{``DCS"}),
\]
and scrambles its challenge as  
\[
C'_j = Ch_j \oplus H(K_{mask} \,\|\, N_j).
\]
It also computes the proof-binding tag  
\[
Proof_j = H(K_{sess} \,\|\, PID_i^{(t)} \,\|\, PID_j^{(t)} \,\|\, N_j).
\]
It then sends  
\[
\{ PID_j^{(t)}, c_j, C'_j, N_j, Proof_j \}
\]
to $D_i$.

\item \textbf{$D_i \rightarrow D_j$: PUF evaluation and proof generation}

Drone $D_i$ recovers the session key via  
\[
K_{sess} = Kyber.Decap(c_j, sk_i^{ep}),
\]
derives $K_{mask}$, and retrieves the honest challenge  
\[
Ch_j = C'_j \oplus H(K_{mask} \,\|\, N_j).
\]
It evaluates its PUF  
\[
Res_i' = PUF(Ch_j),
\]
reconstructs its stable secret  
\[
s_i = FE.Rep(Res_i', w_i),
\]
and computes the hash  
\[
H_i = H(Res_i').
\]
It generates its proof  
\[
Proof_i = H(H_i \,\|\, K_{sess} \,\|\, PID_i^{(t)} \,\|\, N_j),
\]
and sends  
\[
\{ H_i, Proof_i \}
\]
to drone $D_j$.

\item \textbf{$D_j$: Verification and reciprocal PUF proof}

Drone $D_j$ validates $Proof_i$ by recomputing the expected value. It then evaluates its own PUF  
\[
Res_j' = PUF(Ch_i),
\]
computes  
\[
H_j = H(Res_j'),
\]
and constructs its reciprocal proof  
\[
Proof_j' = H(H_j \,\|\, K_{sess} \,\|\, PID_j^{(t)} \,\|\, N_i).
\]
It sends  
\[
\{ H_j, Proof_j' \}
\]
back to drone $D_i$.

\item \textbf{Both drones: Final session key and counter update}

Once mutual proofs are validated, the drones derive the final authenticated session key  
\[
K_{ij} = HKDF(K_{sess} \,\|\, H_i \,\|\, H_j \,\|\, PID_i^{(t)} \,\|\, PID_j^{(t)} \,\|\,
\]
\[
\text{``FINAL"}).
\]
Both then update their counters  
\[
CC_i \leftarrow CC_i + 1,\qquad CC_j \leftarrow CC_j + 1,
\]
ensuring that all future sessions use unlinkable PIDs and DCS masks.

\end{enumerate}


\subsection{Drone Revocation Phase} \label{sec:drone_revo_phase}
\noindent

Drone revocation ensures that a compromised, cloned, or physically captured drone is immediately excluded from all future IoD interactions. Detection of a compromised, cloned, or physically captured drone may occur through various mechanisms. These include detecting anomalous PUF responses, observing authentication failures, or being physically reported by an operator. The freshness and authenticity of revocation broadcasts are ensured through epoch numbers, AEAD protection, and GS-only broadcast authentication using post-quantum digital signatures.

\textbf{Assumptions:}
\begin{enumerate}
    \item Each drone $D_i$ has a long-term identity $ID_i$ bound to its stable PUF-secret $s_i$, and generates unlinkable $PID_i^{(t)}$ in every session.
    
    \item The GS maintains a revocation list $\mathcal{R}$ indexed by $ID_i$ or the corresponding commitment $Com_i$.
    
    \item The GS and all active drones share a symmetric group key $K_{grp}$ derived from Kyber-based GS--drone authentication sessions and periodically refreshed.
    
    \item {Each drone stores the GS public key $pk_{GS}$ for verifying ML-DSA-44 signatures.}
    
    \item AEAD using $K_{grp}$ provides confidentiality and replay protection, while GS signatures ensure that only the GS can issue valid revocation broadcasts.
\end{enumerate}

\textbf{Revocation Procedure:}

\begin{enumerate}

\item When a drone $D_r$ is suspected of compromise (e.g., physical capture, abnormal PUF responses, erratic flight behaviour, or loss of control), the GS inserts its registry entry $(ID_r, Com_r)$ into the global revocation list $\mathcal{R}$.

\item The GS increments the revocation epoch counter $\text{epoch}_{rev}$ and constructs a revocation message,
\[
m_{rev} = \{ID_r, Com_r, \text{epoch}_{rev}\}.
\]

{The GS then signs $m_{rev}$ using the NIST-standardized post-quantum signature scheme ML-DSA-44 \cite{national2024module}, under its long-term signing key $sk_{GS}$,
\[
\sigma_{GS} = \mathrm{ML\text{-}DSA.Sign}_{sk_{GS}}(m_{rev}),
\]}

and encrypts the signed message using AEAD with the group key,
\[
M_{rev} = AEAD_{K_{grp}}(m_{rev}, \sigma_{GS}; AAD = \text{``REVOC''}).
\]

The associated data binds the message to revocation semantics and prevents cross-protocol replay.

\item The GS broadcasts $M_{rev}$ to all drones in the IoD. Although all drones know $K_{grp}$, a malicious or revoked drone cannot forge this broadcast because it does not possess the GS private signing key $sk_{GS}$. Therefore, even if a revoked drone encrypts a fake revocation message using $K_{grp}$, other drones will reject it during signature verification.

\item Each receiving drone $D_i$ performs AEAD decryption using $K_{grp}$. If decryption fails, the message is discarded as unauthentic or tampered.

\item {After successful decryption, drone $D_i$ verifies the GS signature,
\[
\mathrm{ML\text{-}DSA.Verify}_{pk_{GS}}(m_{rev}, \sigma_{GS}).
\]}

If signature verification fails, the message is discarded.
This ensures that only GS-originated revocation messages are accepted.

\item Upon successful verification, $D_i$ extracts $ID_r$ (or $Com_r$) and updates its local revocation cache $\mathcal{R}_i$.
Since drones operate with unlinkable $PID_r^{(t)}$, revocation is applied to the long-term identity $ID_r$ rather than ephemeral PIDs.

\item After updating $\mathcal{R}_i$, drone $D_i$ rejects any future authentication attempt from $D_r$, regardless of the $PID_r^{(t)}$ used.
DCS-scrambled challenges, Kyber ephemeral keys, or authentication attempts originating from $D_r$ are ignored.

\item The revocation remains in effect until the GS performs a future key-refresh epoch.
During the next group key update, a new $K_{grp}'$ is distributed only to non-revoked drones, ensuring that the revoked drone permanently loses access to broadcast communications.

\end{enumerate}

\section{Formal Security Analysis}
\label{sec:formal_security}

In this section, we provide a rigorous formal proof of the security of the proposed post-quantum and PUF-based AKA scheme. We utilize the RoR model, which is the standard for establishing the semantic security and indistinguishability of session keys in the presence of PPT adversaries. Our methodology utilizes the Sequence of Games approach to systematically transform the real protocol execution into an idealized state where the adversary's advantage is provably null.

\subsection{Adversary Model and Oracles}
We adopt the RoR model \cite{bellare1993entity}, assuming an adversary $\mathcal{A}$ that controls the communication medium. Let $\Pi_{D_i}^t$ and $\Pi_{GS}^s$ denote the $t$-th instance of drone $D_i$ and the $s$-th instance of the GS, respectively. We use $\Pi^x$ as a generic placeholder for any protocol instance (drone or GS). The adversary is permitted to execute the following queries:
\begin{itemize}
    \item \textbf{Execute($\Pi_{D_i}^t, \Pi_{GS}^s$):} Models passive eavesdropping on the wireless channel where $\mathcal{A}$ can intercept the messages communicated between Drone $D_i$ and GS.
    \item \textbf{Send($\Pi^x, M$):} Models active attacks, including message modification and injection, where $\mathcal{A}$ can send message $M$ to $\Pi^x$ and then receive a response from it.
    \item \textbf{Reveal($\Pi^x$):} Models the leakage of the session key from a specific instance where $\mathcal{A}$ obtains the session key established between $\Pi^x$ and any other entity.
    \item \textbf{CorruptDrone($D_i$):} Models physical drone capture and extraction of volatile/non-volatile memory.
    \item \textbf{Test($\Pi^x$):} Evaluates the semantic security of the established session key. $\mathcal{A}$ requests $\Pi^x$ for session key $sk$ and receives a probabilistic output based on an unbiased coin or random bit b: (i) if $sk$ is fresh and b = 1, $\Pi^x$ returns $sk$, (ii) if $sk$ is fresh and b = 0, $\Pi^x$ delivers a random number, and (iii) otherwise delivers null.
\end{itemize}

 The RoR model uses Execute, Send,
Reveal, CorruptDevice, and Test queries to simulate an attack
scenario.

\subsection{Formal Security Assumptions}
The security of the framework is predicated on the following cryptographic and physical assumptions.
\begin{enumerate}
    \item \textbf{Uncloneability Assumption:} We define a PUF as $PUF: \{0, 1\}^{l_1} \to \{0, 1\}^{l_2}$, where for a given input of length $l_1$, the PUF outputs a hardware-specific string of length $l_2$. For a challenge-response game consisting of a training and a guess phase, we assume that no PPT adversary $\mathcal{A}$ can guess the response $R_y$ to a fresh challenge $C_y$ (i.e., one not queried during the training phase) \cite{pappu2002physical, gassend2002silicon}. {Thus, the advantage of $\mathcal{A}$, defined as
     \[
    Adv_{\mathcal{A}}^{puf}(l_2) = \Pr\left[ R'_y = R_y \right],
    \] is negligible in $l_2$. Here, $R'_y$ is the response output by $\mathcal{A}$ in the guess phase, $R_y = PUF(C_y)$, and the probability is taken over all possible PUF instances and adversary strategies.
    }
    \item \textbf{Pseudo-random Function (PRF) Assumption:} Assume that $h(\cdot)$ is a polynomial-time computable hash function. To distinguish $h(\cdot)$ from a random function, an adversary $\mathcal{A}$ may request polynomially bounded queries with its selected inputs and obtain the outputs computed by $h(\cdot)$ for training. After completion of the training phase, $\mathcal{A}$ is provided a function, which is either $h(\cdot)$ or a truly random function. We say that $h(\cdot)$ is a $\epsilon$-secure PRF if it is indistinguishable from a random function under $\mathcal{A}$ such that $\mathcal{A}$ can only distinguish the two with a probability of $1/2 + \epsilon$.
    \item \textbf{MLWE Hardness:} The security of Kyber KEM relies on the computational difficulty of the MLWE problem. We assume that for any PPT adversary, the advantage $Adv^{MLWE}(\mathcal{A})$ in distinguishing the public Kyber matrix $\mathbf{A}$ and vector $\mathbf{t}$ (see Section \ref{SSC:MLWE}) from uniform samples over the ring $R_q$ is negligible.
\end{enumerate}

\subsection{Security Bound}\label{sec:formal_sec_theorem_proof}
\begin{theorem}\label{theorem1_statement}
Let $\mathcal{A}$ be a PPT adversary running in time $t$ against the proposed protocol $\Pi$. The advantage $Adv_{\mathcal{A}}^{AKE}(t)$ of $\mathcal{A}$ in compromising the semantic security of the session key $SK_{is}$ is bounded as follows:
\begin{equation}
\begin{split}
Adv_{\mathcal{A}}^{AKE}(t) \leq\;& \frac{q_h^2}{2|Hash|} 
+ \frac{(q_s + q_e)^2}{2^{|N|}} + 2 \cdot Adv^{MLWE}(\mathcal{A}) \\
& + \frac{q_p^2}{2|PUF|} + \frac{2q_{send}}{|D|},
\end{split}
\label{EQ:advantage:bound}
\end{equation}
where $q_h, q_p, q_{send}$ are the number of Hash, PUF, and Send queries, respectively, {$q_s$ and $q_e$ are the number of sessions initiated by $\mathcal{A}$ through Send and Execute queries, respectively,} and $|N|$ denotes the bit length of the nonces, $|Hash|$ and $|PUF|$ denote the sizes of the output spaces of the hash function and the PUF, respectively, and $|D|$ denotes the size of the drone set $D$.
\end{theorem}
\begin{IEEEproof}\label{theorem_proof}
The proof proceeds through a sequence of seven games, $\mathcal{G}_0, \mathcal{G}_1, \dots, \mathcal{G}_6$. Let $Succ_{\mathcal{A}}^{\mathcal{G}_i}$ denote the event that $\mathcal{A}$ correctly guesses the random bit $b$ in Game $\mathcal{G}_i$.  The advantage of the adversary to win the game can then be defined as 

\[
Adv_{\mathcal{A}, \mathcal{G}_i}^{AKE}(t) = \left| 2\Pr[Succ_{\mathcal{A}}^{\mathcal{G}_i}] - 1 \right|.
\]

\textbf{Game $\mathcal{G}_0$:} This game corresponds to the real attack scenario in the RoR model. The simulator executes the protocol using real parameters for Kyber, PUF responses, and hash values. The advantage of $\mathcal{A}$ is defined as
\begin{equation} 
   Adv^{AKE}_{\mathcal{A}}(t) = Adv_{\mathcal{A}, \mathcal{G}_0}^{AKE}(t) = |2Pr[Succ_{\mathcal{A}}^{\mathcal{G}_0}] - 1|. \label{EQ:G0}
\end{equation}
This is the baseline definition of the adversary's advantage in the RoR model. It measures how well an adversary \(\mathcal{A}\) can distinguish a real session key from a randomly generated one.

An advantage of \(0\) indicates that \(\mathcal{A}\) cannot distinguish between the two cases, implying perfect indistinguishability and therefore full security. Conversely, an advantage of \(1\) indicates that \(\mathcal{A}\) can always distinguish correctly, implying complete compromise of the protocol.

\textbf{Game $\mathcal{G}_1$:} $\mathcal{A}$ attempts an eavesdropping attack by conducting an Execute query. Further, $\mathcal{A}$ runs Test queries to determine if the acquired value is a session key or not. $\mathcal{A}$ must know
$K_{sess}, R'_i, PID_i^{(t)}$ to acquire the session key $SK_{is} = HKDF(K_{sess} \,\|\, R'_i \,\|\, PID_i^{(t)})$. However, these values cannot be obtained by eavesdropping attacks. This means that $\mathcal{A}$ has no advantage to be gained through an Execute query. Therefore, the probability of $\mathcal{A}$
winning $\mathcal{G}_1$ is equal to that of $\mathcal{A}$ winning $\mathcal{G}_0$. 
\begin{equation}
    Pr[Succ_{\mathcal{A}}^{\mathcal{G}_1}] = Pr[Succ_{\mathcal{A}}^{\mathcal{G}_0}]. \label{EQ:G1:G0:diff}
\end{equation}

\textbf{Game $\mathcal{G}_2$:} This game simulates an active attack by hash
queries. We model all cryptographic hash and HKDF functions as random oracles. In this game, $\mathcal{A}$ runs $\textsf{Send}$ and \textsf{Hash} queries to expose the session key. The
transmitted messages can be modified by $\mathcal{A}$. However, $\mathcal{A}$ should find a hash collision to win the game because all transmitted messages are masked by a one-way function $H(\cdot)$ as well as the DCS algorithm. Therefore, the advantage that $\mathcal{A}$ can gain at the end of $\mathcal{G}_2$ is obtained based on the
birthday paradox \cite{bellare1993random,katz2020introduction}, considering the probability of a hash collision during $q_h$ queries, bounded as follows: 
\begin{equation}
    |Pr[Succ_{\mathcal{A}}^{\mathcal{G}_2}] - Pr[Succ_{\mathcal{A}}^{\mathcal{G}_1}]| \leq \frac{q_h^2}{2|Hash|}. \label{EQ:G2:G1:diff}
\end{equation}

\textbf{Game $\mathcal{G}_3$:} The simulator ensures the uniqueness of all ephemeral nonces $N$ and monotonic counters $CC_i$. If a collision occurs in the nonce space $\{0, 1\}^{128}$, the simulation is aborted. This is similar to the game $\mathcal{G}_2$. The probability that any two of the \(q_s + q_e\) independently generated 128-bit nonces collide is bounded by the birthday bound as
$\frac{(q_s + q_e)^2}{2^{129}}$, where the additional factor of \(2\) in the denominator accounts for the abort condition in the reduction. Since $|N| = 128$,
the collision probability is negligible. For $q_s + q_e$ sessions, the difference between  success probabilities is bounded as follows:
\begin{equation}
    |Pr[Succ_{\mathcal{A}}^{\mathcal{G}_3}] - Pr[Succ_{\mathcal{A}}^{\mathcal{G}_2}]| \leq \frac{(q_s + q_e)^2}{2^{|N|+1}}. \label{EQ:G3:G2:diff}
\end{equation}

\textbf{Game $\mathcal{G}_4$:} We address the post-quantum key exchange. $\mathcal{A}$ runs the Send query. The simulator replaces the Kyber ciphertext $c_{GS}$ and shared secret $K_{sess}$ with random values. Any adversary capable of distinguishing this game from $\mathcal{G}_3$ can solve the MLWE problem. The factor of 2 arises because MLWE security is used twice in the Kyber construction (for the public key and the ciphertext). The success difference is thus bounded by the MLWE advantage:
\begin{equation}
    |Pr[Succ_{\mathcal{A}}^{\mathcal{G}_4}] - Pr[Succ_{\mathcal{A}}^{\mathcal{G}_3}]| \leq 2 \cdot Adv^{MLWE}(\mathcal{A}). \label{EQ:G4:G3:diff}
\end{equation}

\textbf{Game $\mathcal{G}_5$:} This game targets the DCS mechanism and PUF response. $\mathcal{A}$ runs Send and PUF queries. The scrambled challenge $C'_i$ and proof $Proof_i$ are replaced with random values. Since $K_{mask}$ is derived from $K_{sess}$, which was randomized in $\mathcal{G}_4$, the mask is unknown to $\mathcal{A}$. Thus, $\mathcal{A}$ obtains no advantage from game $\mathcal{G}_4$. By the PUF unpredictability assumption (Section \ref{subsec:queries}), the probability of correctly guessing any PUF response across \(q_p\) queries is bounded by $\frac{q_p^2}{2|PUF|}$,
which is analogous to the birthday-bound argument used for the hash function in \(\mathcal{G}_2\).
\begin{equation}
    |Pr[Succ_{\mathcal{A}}^{\mathcal{G}_5}] - Pr[Succ_{\mathcal{A}}^{\mathcal{G}_4}]|
 \leq \frac{q_p^2}{2|PUF|} \label{EQ:G5:G4:diff}
 \end{equation}

\textbf{Game $\mathcal{G}_6$:} In this game, $\mathcal{A}$ performs Corrupt and Send queries to obtain all the information stored in drones and attempts to impersonate a legitimate drone to establish a valid session.

The simulator assumes that $\mathcal{A}$ tries to guess the correct drone identity and corresponding session to derive the session key $SK_{is}$. Since the drone identity $ID_i$ and its corresponding session parameters are selected uniformly from the drone set $D$, the probability that $\mathcal{A}$ successfully selects the correct target drone during $q_{send}$ Send queries is bounded by
\[
\frac{q_{send}}{|D|}.
\]

Because the RoR model considers both sides of the guessing advantage and applies triangle inequality over game transitions, the total difference between the success probabilities of $\mathcal{G}_6$ and $\mathcal{G}_5$ is bounded as follows:
\begin{equation}
|Pr[Succ_{\mathcal{A}}^{\mathcal{G}_6}] - Pr[Succ_{\mathcal{A}}^{\mathcal{G}_5}]|
\leq \frac{2q_{send}}{|D|}.
\label{EQ:G6:G5:diff}
\end{equation}

In this final game, all cryptographic elements including Kyber shared secrets, PUF responses, and hash outputs are replaced with random values. Therefore, the adversary has no information about the session key and can only guess the hidden bit $b$ with probability
\[
Pr[Succ_{\mathcal{A}}^{\mathcal{G}_6}] = \frac{1}{2}.
\]

Applying the triangle inequality over all game transitions, we obtain
\begin{align}
Adv^{AKE}_{\mathcal{A}}(t)
&=
\left|
2 \Pr\!\left[\mathrm{Succ}^{\mathcal{G}_0}_{\mathcal{A}}\right] - 1
\right|
\\[4pt]
&\leq
\left|
2 \Pr\!\left[\mathrm{Succ}^{\mathcal{G}_6}_{\mathcal{A}}\right] - 1
\right|
\nonumber\\
&\quad+
\sum_{i=0}^{5}
\left|
\Pr\!\left[\mathrm{Succ}^{\mathcal{G}_{i+1}}_{\mathcal{A}}\right]
-
\Pr\!\left[\mathrm{Succ}^{\mathcal{G}_i}_{\mathcal{A}}\right]
\right|
\\[6pt]
&=
0
+
\frac{q_h^2}{2|Hash|}
+
\frac{(q_s + q_e)^2}{2^{|N|}}
\nonumber\\
&\quad+
2 \mathrm{Adv}^{MLWE}(\mathcal{A})
+
\frac{q_p^2}{2|PUF|}
+
\frac{2 q_{send}}{|D|}.
\end{align}

The above inequality follows by applying:
\begin{itemize}
    \item Equation~(4) for the transition \(\mathcal{G}_0 \rightarrow \mathcal{G}_1\),
    \item Equation~(5) for the transition \(\mathcal{G}_1 \rightarrow \mathcal{G}_2\),
    \item Equation~(6) for the transition \(\mathcal{G}_2 \rightarrow \mathcal{G}_3\),
    \item Equation~(7) for the transition \(\mathcal{G}_3 \rightarrow \mathcal{G}_4\),
    \item Equation~(8) for the transition \(\mathcal{G}_4 \rightarrow \mathcal{G}_5\),
    \item Equation \eqref{EQ:G6:G5:diff} for the transition \(\mathcal{G}_5 \rightarrow \mathcal{G}_6\).
\end{itemize}
Also, since
\[
\Pr\!\left[\mathrm{Succ}^{\mathcal{G}_6}_{\mathcal{A}}\right] = \frac{1}{2},
\]
the adversary in \(\mathcal{G}_6\) can do no better than random guessing. Therefore,
\[
\left|
2 \Pr\!\left[\mathrm{Succ}^{\mathcal{G}_6}_{\mathcal{A}}\right] - 1
\right|=\left|
2 \cdot \frac{1}{2} - 1
\right|
= 0,
\]
and the bound given in \eqref{EQ:advantage:bound} follows.
\end{IEEEproof}

\section{ProVerif-Based Formal Verification}
\label{sec:proverif}

To further validate the security of the proposed protocol, we performed
a formal verification using the ProVerif tool~\cite{blanchet2016modeling}.
ProVerif is an automated cryptographic protocol verifier based on the applied pi-calculus ~\cite{abadi2001mobile} and is widely used for
mechanically proving properties such as secrecy, authentication, and correspondence in security protocols operating over insecure channels.
It models the adversary as a Dolev--Yao attacker
who has full control over the communication channel, including the ability to intercept, modify, replay, and inject messages.
Unlike simulation-based tools, ProVerif provides proofs that hold for an unbounded number of sessions, making it suitable for demonstrating strong security guarantees.

Our protocol involves several operations such as Kyber KEM encapsulation and decapsulation, PUF evaluation, fuzzy extractor reproduction, DCS-based challenge scrambling, and AEAD authentication, which are not natively supported in ProVerif's algebraic model. We introduce a set of protocol simplifications and abstractions that preserve the core security structure of the original design.
Section~\ref{subsec:assumptions} formally states these assumptions and provides justification for their soundness.

\subsection{Assumptions and Justification}
\label{subsec:assumptions}

\subsubsection{Kyber KEM Modeled as an Ideal KEM}

The Kyber Key Encapsulation Mechanism is modeled as an ideal,
perfectly binding KEM in which the encapsulation function
\texttt{KyberEncap}$(pk)$ produces a ciphertext, and the shared secret
is recovered via \texttt{KyberDecap}$(ct, sk)$, with the correctness
equation
\[
\begin{aligned}
\texttt{KyberDecap}(
    \texttt{KyberEncap}(pk),\; sk
)
&=
\\
& \texttt{KyberShared}(pk).
\end{aligned}
\]
This is a standard abstraction \cite{blanchet2016modeling} that treats the KEM as a black box.

\subsubsection{PUF Modeled as a Private, Uncloneable Function}

The PUF is declared with the
\texttt{[private]} attribute in ProVerif, meaning the adversary
cannot evaluate or invert it.
Formally, \texttt{PUF} is modeled as a free function
$\texttt{PUF} : \texttt{pufChallenge} \to \texttt{pufResponse}$
that is inaccessible to the attacker.
This accurately captures the PUF unpredictability assumption
(Section \ref{sec:PUF}).

\subsubsection{Fuzzy Extractor Modeled via a Rewrite Rule}

The fuzzy extractor's \textit{reproduce} function
\texttt{FE.Rep} is modeled using a ProVerif \texttt{reduc} rewrite
rule:
\[
\begin{aligned}
\forall r : \texttt{pufResponse},\quad
&\texttt{FErep}(r,\; \texttt{FEhelper}(r))
\\
&= \texttt{FEgen}(r).
\end{aligned}
\]
This equation captures the correctness property, which says that given the helper data $w_i = \texttt{FEhelper}(\mathit{Res}_i)$ and
any fresh PUF response $\mathit{Res}'_i$ evaluated on the same
challenge, \texttt{FE.Rep} reproduces the same stable secret $s_i$.

\subsubsection{AEAD Modeled as Symmetric Authenticated Encryption}

The AEAD primitive (ChaCha20-Poly1305 in the concrete scheme) is
modeled as a symmetric authenticated encryption scheme using a
ProVerif \texttt{reduc} rule that captures the decryption correctness
equation $\texttt{AEADdec}(k, \texttt{AEADenc}(k, m, \mathit{aad}),
\mathit{aad}) = m$.

\subsubsection{Hash Functions and HKDF Modeled as Free Functions}

The hash function $H(\cdot)$ and key derivation function
$\mathrm{HKDF}(\cdot)$ are modeled as free, injective functions over
the ProVerif symbolic term algebra.
This corresponds to modeling them as random oracles, which is the
standard assumption used in the security proof (Section \ref{sec:formal_security}).
In the symbolic model, the adversary cannot invert these functions
or find collisions.

\subsubsection{DCS Modeled as a Keyed Scrambling Function}

The DCS operation
$C'_i = Ch_i \oplus H(K_{\mathrm{mask}} \| N_{GS})$ is modeled as a
free function $\texttt{DCS}(Ch_i, K_{\mathrm{mask}}, N_{GS})$.
Since $K_{\mathrm{mask}}$ is derived from the Kyber session key
$K_{\mathrm{sess}}$ (which is private to the participants), the
adversary cannot compute the DCS output or invert it to obtain the
true challenge $Ch_i$.
This correctly models the session-obfuscation property of DCS.

\subsection{ProVerif Queries}
\label{subsec:queries}

Three security queries are posed to ProVerif, corresponding to the three
key properties of the proposed protocol:

\begin{enumerate}
  \item[\textit{Q1.}] \textit{Session key secrecy:}
        The established session key $SK_{is}$ must be secret from the
        adversary in all reachable states.
  \item[\textit{Q2.}] \textit{PUF stable secret secrecy:}
        The stable secret $s_i$ derived from the PUF response must
        remain secret from the adversary.
  \item[\textit{Q3.}] \textit{Drone-to-GS authentication
        (injective correspondence):} Whenever the GS completes
        authentication of a drone, the drone must have previously
        initiated that session.
        This is an injective correspondence query, ensuring
        that replays are detected.
\end{enumerate}


\subsection{Protocol Implementation}
Listing~\ref{lst:types_channels} shows the type declarations and
channel definitions.
Two channels are declared: \texttt{secChan}, which models the secure
provisioning channel used only during enrollment, and
\texttt{pubChan}, which models the insecure public wireless channel
over which all subsequent authentication traffic flows.

\begin{lstlisting}[language=ProVerif,
  caption={Type declarations and channel definitions.},
  label={lst:types_channels}]
(* ----- Type Declarations ----- *)
type droneID.
type pseudoID.
type kemPubKey.
type kemSecKey.
type sessionKey.
type pufChallenge.
type pufResponse.
type helperData.
type commitment.

(* ----- Channels ----- *)
(* Secure provisioning channel: used only during enrollment *)
channel secChan.
(* Insecure public wireless channel: all authentication traffic *)
channel pubChan.
\end{lstlisting}

Listing~\ref{lst:functions} defines all cryptographic functions and
rewrite rules.
The Kyber KEM is modeled using \texttt{KyberEncap} (encapsulation),
\texttt{KyberShared} (shared secret), and \texttt{KyberDecap}
(decapsulation), linked by an \texttt{equation} that enforces the
KEM correctness property.
The PUF function and the fuzzy extractor's generate function
\texttt{FEgen} are declared \texttt{[private]}, ensuring the adversary
cannot evaluate them.
The \texttt{reduc} rule for \texttt{FErep} captures the
error-correction property of the fuzzy extractor.

\begin{lstlisting}[language=ProVerif,
  caption={Cryptographic function declarations and rewrite rules.},
  label={lst:functions}]
(* Hash function (SHA-256), modeled as a random oracle *)
fun H(bitstring): bitstring.

(* HKDF key derivation *)
fun HKDF2(sessionKey, bitstring): sessionKey.
fun HKDFbits(sessionKey, bitstring): bitstring.

(* Kyber KEM: ideal black-box model *)
fun KyberEncap(kemPubKey): bitstring.       (* produces ciphertext *)
fun KyberShared(kemPubKey): sessionKey.     (* produces Ksess      *)
fun KyberDecap(bitstring, kemSecKey): sessionKey.

equation forall epk: kemPubKey, esk: kemSecKey;
  KyberDecap(KyberEncap(epk), esk) = KyberShared(epk).

(* AEAD authenticated encryption *)
fun AEADenc(sessionKey, bitstring, bitstring): bitstring.
reduc forall k: sessionKey, m: bitstring, aad: bitstring;
  AEADdec(k, AEADenc(k, m, aad), aad) = m.

(* PUF: private - adversary cannot evaluate *)
fun PUF(pufChallenge): pufResponse [private].

(* Fuzzy extractor *)
fun FEgen(pufResponse): bitstring [private].   (* generate: private *)
fun FEhelper(pufResponse): helperData.         (* helper data: public *)
reduc forall r: pufResponse;
  FErep(r, FEhelper(r)) = FEgen(r).            (* reproduce equation *)

(* Pseudo-identity: H(IDi || CCi || "PID") *)
fun PseudoID(droneID, bitstring): pseudoID.

(* Commitment: H(IDi || pki || si) *)
fun Commit(droneID, kemPubKey, bitstring): commitment.

(* DCS scrambling: Chi XOR H(Kmask || NGS) *)
fun DCS(pufChallenge, bitstring, bitstring): pufChallenge.

(* Authentication proof: H(R'i || Ksess || NGS || PIDi) *)
fun MkProof(pufResponse, sessionKey, bitstring, pseudoID): bitstring.

(* Final session key derivation: HKDF(Ksess || R'i || PIDi) *)
fun DeriveSK(sessionKey, pufResponse, pseudoID): sessionKey.
\end{lstlisting}

Listing~\ref{lst:events_queries} declares the protocol events and
the four security queries.
Events \texttt{DroneBegins} and \texttt{GSCompletesAuth} are used to
express the correspondence for drone authentication (Q3), while
\texttt{GSCompletesAuth} and \texttt{DroneCompletesAuth} are used for
GS authentication (Q4).
The secrecy queries (Q1, Q2) use the \texttt{attacker} predicate,
which is true if the adversary can derive the term.
The correspondence queries use \texttt{inj-event} to enforce
injectivity.

\begin{lstlisting}[language=ProVerif,
  caption={Event declarations and security queries.},
  label={lst:events_queries}]
(* ----- Events ----- *)
event DroneBegins(droneID, pseudoID, kemPubKey).
event GSCompletesAuth(droneID, pseudoID, sessionKey).
event DroneCompletesAuth(droneID, pseudoID, sessionKey).
event SessionEstablished(droneID, sessionKey).

(* ----- Queries ----- *)

(* Q1: Session key secrecy *)
query sk: sessionKey; attacker(sk).

(* Q2: PUF-derived stable secret secrecy *)
query si: bitstring; attacker(si).

(* Q3: Injective correspondence - drone-to-GS authentication *)
query IDi: droneID, PIDi: pseudoID, SK: sessionKey;
  inj-event(GSCompletesAuth(IDi, PIDi, SK)) ==>
  inj-event(DroneBegins(IDi, PIDi, new kemPubKey)).

(* Q4: Injective correspondence - GS-to-drone authentication *)
query IDi: droneID, PIDi: pseudoID, SK: sessionKey;
  inj-event(DroneCompletesAuth(IDi, PIDi, SK)) ==>
  inj-event(GSCompletesAuth(IDi, PIDi, SK)).
\end{lstlisting}

Listing~\ref{lst:drone_process} defines the drone process.
The drone first derives its session pseudo-identity
$\mathit{PID}^{(t)}_i = H(ID_i \| CC_i \| \text{``PID''})$, generates
an ephemeral Kyber key pair, and sends both to the GS on the public
channel.
Upon receiving the GS ciphertext $c_{GS}$ and nonce $N_{GS}$, the
drone decapsulates to recover $K_{\mathrm{sess}}$, computes the DCS
mask, evaluates the PUF on the scrambled challenge, reproduces $s_i$
via the fuzzy extractor, and constructs the proof
$\mathit{Proof}_i$.
After receiving the AEAD-protected acknowledgment and verifying it,
the drone signals \texttt{DroneCompletesAuth} to record successful GS
authentication.

\begin{lstlisting}[language=ProVerif,
  caption={Drone process definition.},
  label={lst:drone_process}]
let Drone(IDi: droneID,
          Chi: pufChallenge,
          wi: helperData,
          pkGS: kemPubKey,
          CCi: bitstring,
          pkEph: kemPubKey,
          skEph: kemSecKey) =
  (* Step 1: Derive session pseudo-identity *)
  let PIDi = PseudoID(IDi, CCi) in
  (* Step 2: Send PID and ephemeral public key to GS *)
  event DroneBegins(IDi, PIDi, pkEph);
  out(pubChan, (PIDi, pkEph));
  (* Step 3: Receive Kyber ciphertext and GS nonce *)
  in(pubChan, (cGS: bitstring, NGS: bitstring));
  (* Step 4: Decapsulate to recover session key *)
  let Ksess = KyberDecap(cGS, skEph) in
  (* Step 5: Derive DCS mask and scramble enrollment challenge *)
  let Kmask = HKDFbits(Ksess, "DCS") in
  let Cprime = DCS(Chi, Kmask, NGS) in
  (* Step 6: Evaluate PUF and reproduce stable secret *)
  let Rprime = PUF(Cprime) in
  let si_rep = FErep(Rprime, wi) in
  (* Step 7: Compute authentication proof *)
  let Proofi = MkProof(Rprime, Ksess, NGS, PIDi) in
  (* Step 8: Send scrambled challenge and proof *)
  out(pubChan, (Cprime, Proofi));
  (* Step 9: Receive and verify GS acknowledgment *)
  in(pubChan, encAck: bitstring);
  let ackKey = HKDF2(Ksess, "ACK") in
  let ack = AEADdec(ackKey, encAck, "AUTH-ACK") in
  (* Step 10: Derive final session key and record completion *)
  let SK = DeriveSK(Ksess, Rprime, PIDi) in
  event DroneCompletesAuth(IDi, PIDi, SK);
  event SessionEstablished(IDi, SK).
\end{lstlisting}

Listing~\ref{lst:gs_process} defines the GS process.
The GS receives the drone's pseudo-identity and ephemeral public key,
performs Kyber encapsulation to produce the session key and ciphertext,
and sends both back together with a fresh nonce $N_{GS}$.
Upon receiving the scrambled challenge $C'_i$ and proof
$\mathit{Proof}_i$, the GS recomputes the expected proof using the
stored commitment and verifies both.
If both checks pass, the GS fires the \texttt{GSCompletesAuth} event
and sends an AEAD-protected acknowledgment, completing the authentication
from the GS side.

\begin{lstlisting}[language=ProVerif,
  caption={Ground Station process definition.},
  label={lst:gs_process}]
let GS(IDi: droneID,
       Comi: commitment,
       pkiStored: kemPubKey,
       siStored: bitstring,
       Chi: pufChallenge) =
  (* Step 1: Receive drone PID and ephemeral public key *)
  in(pubChan, (PIDi: pseudoID, pkEph: kemPubKey));
  (* Step 2: Perform Kyber encapsulation *)
  let cGS = KyberEncap(pkEph) in
  let Ksess = KyberShared(pkEph) in
  new NGS: bitstring;
  out(pubChan, (cGS, NGS));
  (* Step 3: Receive scrambled challenge and proof *)
  in(pubChan, (Cprime: pufChallenge, Proofi: bitstring));
  (* Step 4: Recompute expected proof using stored parameters *)
  let Rprime = PUF(Cprime) in
  let expectedProof = MkProof(Rprime, Ksess, NGS, PIDi) in
  (* Step 5: Verify proof and commitment binding *)
  if Proofi = expectedProof then
  if Commit(IDi, pkiStored, siStored) = Comi then
  (* Step 6: Authentication successful - fire event *)
  event GSCompletesAuth(IDi, PIDi, Ksess);
  let SK = DeriveSK(Ksess, Rprime, PIDi) in
  (* Step 7: Send AEAD-protected acknowledgment *)
  let ackKey = HKDF2(Ksess, "ACK") in
  let encAck = AEADenc(ackKey, "SUCCESS", "AUTH-ACK") in
  out(pubChan, encAck).
\end{lstlisting}

Listing~\ref{lst:main} defines the main process.
The GS long-term key pair is generated at the start.
Enrollment is performed once: the drone's PUF is evaluated on the
assigned challenge $Ch_i$, the stable secret $s_i$ and helper data
$w_i$ are derived via the fuzzy extractor, and the commitment
$\mathit{Com}_i$ is computed and stored.
The replication operator \texttt{!} causes the drone and GS processes
to run concurrently for an unbounded number of sessions with fresh
ephemeral key pairs, which models the long-term deployment scenario
and allows ProVerif to reason about all possible session interleavings.

\begin{lstlisting}[language=ProVerif,
  caption={Main process: enrollment and unbounded authentication sessions.},
  label={lst:main}]
process
  (* Long-term GS key pair *)
  new pkGS: kemPubKey;
  new skGS: kemSecKey;
  (* One-time drone enrollment *)
  new IDi: droneID;
  new Chi: pufChallenge;
  new CCi: bitstring;
  let Resi = PUF(Chi) in
  let si = FEgen(Resi) in
  let wi = FEhelper(Resi) in
  new pkiLT: kemPubKey;
  let Comi = Commit(IDi, pkiLT, si) in
  (* Unbounded authentication sessions with fresh ephemeral keys *)
  !(
    new pkEph: kemPubKey;
    new skEph: kemSecKey;
    ( Drone(IDi, Chi, wi, pkGS, CCi, pkEph, skEph)
    | GS(IDi, Comi, pkiLT, si, Chi) )
  )
\end{lstlisting}

\subsection{Verification Results}
\label{subsec:results}

\begin{figure}[!htbp]
  \centering
  %
  \includegraphics[width=1\linewidth]{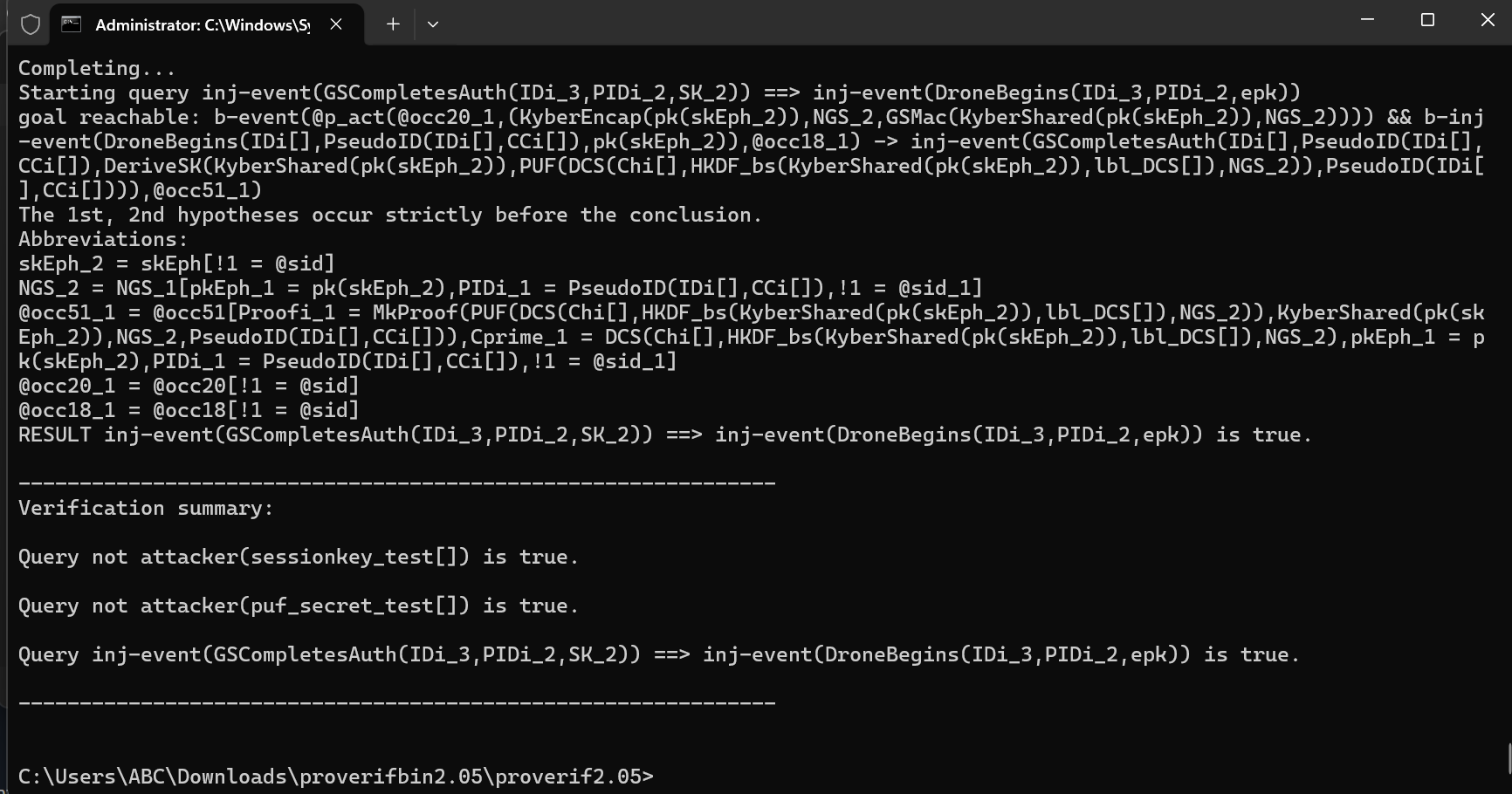}
  
  \caption{ProVerif~2.05 terminal output showing successful verification
           of three security properties.
           All queries return \texttt{true}, confirming that the proposed
           protocol is \textsc{safe} under the Dolev--Yao adversary model.}
  \label{fig:proverif_result}
\end{figure}

The protocol was verified using ProVerif~2.05 on a machine running
Windows 10 with an Intel Core i3 processor.
The complete verification, covering all three queries and an unbounded
number of sessions, completed in under one second.
Fig.~\ref{fig:proverif_result} shows a screenshot of the ProVerif
terminal output.
The verification results are summarized below.

Table~\ref{tab:proverif_summary} summarizes the three verified security
properties, the query type used, and the verification result.

\begin{table}[!htbp]
  \centering
  \caption{Summary of ProVerif Verification Results.}
  \label{tab:proverif_summary}
  \begin{tabular}{|l|l|c|c|}
    \hline
    \textbf{Property} & \textbf{Query Type} & \textbf{Sessions} & \textbf{Result} \\
    \hline
    Session key secrecy      & \texttt{attacker()} & Unbounded & \textbf{true} \\
    PUF stable secret secrecy & \texttt{attacker()} & Unbounded & \textbf{true} \\
    Drone-to-GS authentication & \texttt{inj-event()} & Unbounded & \textbf{true} \\
    \hline
  \end{tabular}
\end{table}

The result \texttt{true} for Q1 confirms that the established session
key $SK_{is}$ cannot be derived by the Dolev--Yao adversary in any
reachable state, regardless of the number of protocol sessions
observed or actively manipulated.
This result holds because the adversary cannot decapsulate the Kyber
ciphertext $c_{GS}$ without the drone's ephemeral secret key
$sk^{ep}_i$ (modeled via the ideal KEM equation), and therefore cannot
derive $K_{\mathrm{sess}}$ or any value derived from it.

The result \texttt{true} for Q2 confirms that the PUF-derived stable
secret $s_i$ remains confidential.
This holds because \texttt{PUF} and \texttt{FEgen} are both declared
\texttt{[private]}, making them inaccessible to the adversary even if
the helper data $w_i$ or the scrambled challenge $C'_i$ are observed
on the public channel.

The \texttt{inj-event} result \texttt{true} for Q3 confirms
\emph{injective} agreement: whenever the GS fires
\texttt{GSCompletesAuth(IDi, PIDi, SK)}, there is a distinct,
matching instance in which the drone fired
\texttt{DroneBegins(IDi, PIDi, pkEph)} for the same session.
Injectivity ensures that the GS cannot be fooled by a replay of a
previous session's transcript, which directly confirms replay-attack
resistance.

\subsection{Discussion}
\label{subsec:proverif_discussion}

The ProVerif verification complements the sequence-of-games proof in
Section \ref{sec:formal_security} by providing a mechanically-checked,
model-theoretic confirmation of the protocol's security structure.
The ProVerif analysis establishes symbolic
security under the Dolev--Yao model for an unbounded number of
sessions, verifying that the protocol logic itself is sound and that
no attack trace exists in the symbolic model.

\section{Informal Security Analysis}\label{sec:informal_sec_proof}
\subsection{Replay Attack}
Replay attacks are prevented through the use of per-session nonces and Kyber-generated ephemeral secrets. Every authentication round incorporates fresh nonces $N_i$ and $N_{GS}$, which are mixed directly into $K_{sess}$, the DCS mask $Mask_i$, and all AEAD associated data. Because $PID_i^{(t)}$ also evolves with the monotonic counter, an attacker cannot replay an older tuple $\{PID_i^{(t)}, c_{GS}, C'_i, Proof_i\}$ without immediate failure.

\subsection{Impersonation Attack}
Our scheme is impersonation resistant because of various aspects. First, the attacker cannot forge Kyber-based session keys since decapsulation requires the secret key of the victim drone, which is sealed behind its PUF and cannot be extracted. Second, even if the attacker guesses or reuses an old challenge, the DCS ensures that $C'_i$ changes unpredictably every session through $Mask_i = H(K_{mask} \| N_{GS})$. Without access to $K_{sess}$, the attacker cannot compute the correct scrambled challenge, nor can they compute $Proof_i = H(R'_i \| K_{sess} \| N_{GS} \| PID_i^{(t)})$. Also, because the GS stores only a one-way commitment $Com_i$, an attacker cannot derive the required PUF response from any stolen backend data.

\subsection{Drone Physical Capture Attack}

{If a drone is physically captured, the adversary can extract non-critical stored material such as the enrollment challenge $Ch_i$, helper data $w_i$, and the monotonic counter $CC_i$. Since the attacker now possesses the genuine drone hardware, they can perform live PUF evaluations using the extracted $Ch_i$ and $w_i$ and thereby fully recover $s_i$ via $FE.Rep(PUF_i(Ch_i),w_i)$. However, this compromise does not extend beyond the captured UAV. Since $s_i$ is a hardware-bound PUF output that cannot be extracted or transferred without the real hardware, knowledge of $s_i$ for Drone $D_i$ provides no information about $s_j$ for any other drone $D_j$. Furthermore, once the compromise is detected (Section \ref{sec:drone_revo_phase}), the GS adds ($ID_i$, $Com_i$) to the revocation list, and all peer drones subsequently reject $D_i$'s pseudo-identities $PID^{(t)}_i$, permanently excluding the captured drone from future sessions. Probing or invasive access to the PUF hardware also destroys its analog structure, permanently preventing extraction of $s_i$ \cite{10.1145/2818186}. }

\subsection{GS Physical Capture Attack}

{GS compromise does not leak the drone secret $s_i$ or allow derivation of session keys. This is because the GS never stores raw PUF responses, never stores $s_i$, and does not hold any drone secret keys. The only per-drone value maintained by the GS is a one-way commitment $Com_i = H(ID_i \| pk_i \| s_i)$, which is useless to an attacker even under full registry disclosure. This guarantee, however, is limited to static GS data. Since the GS serves as the sole registry, directory, and revocation authority (Section \ref{sec:entities_involved}), an adversary that gains persistent operational control of the live GS process could disrupt the GS's AKA sessions and deny services to all drones. Addressing such attacks remains an important direction for future research.}

\subsection{Forward and Backward Secrecy}
The use of Kyber KEM for every authentication provides forward and backward secrecy \cite{alagic2024module}. Forward secrecy is ensured because each session generates a fresh $K_{sess}$ derived from randomized Kyber encapsulation. Even if an adversary later compromises a drone and retrieves its current private key, they cannot retroactively decrypt old traffic or reconstruct earlier session keys.

{Similarly, backward secrecy is also achieved in the proposed architecture. This is because, even if an adversary obtains the current session key  at time $t$, future session keys at time $t+1$ remain secure. Each drone-GS or D2D session generates a fresh ephemeral Kyber key pair using independent randomness. This is precisely the argument used in the transition to Game $\mathcal{G}_4$ of the RoR proof (Section \ref{theorem_proof}), where $K_{sess}$ was replaced by a value indistinguishable from random under the MLWE hardness assumption.}

\subsection{Man-in-the-Middle (MITM) Attack}
A MITM adversary can intercept, delay, reorder, and inject arbitrary messages, but cannot derive or modify $K_{sess}$. Kyber encapsulation ensures that only the legitimate holder of $sk_i^{ep}$ can decapsulate $c_{GS}$, so a MITM cannot compute $K_{sess}$ and therefore cannot compute the correct DCS mask or decrypt/alter AEAD-protected data.

\subsection{Desynchronisation Attack}
The proposed scheme is explicitly designed to avoid reliance on timestamps, thus eliminating the primary cause of desynchronization in UAV networks. Freshness is instead guaranteed by nonces and per-drone monotonic counters.

\subsection{Insider Attack}
{\textbf{Insider capability assumptions:} We model an insider adversary as a legitimately enrolled, but malicious drone $D_m$ that follows the protocol honestly, but tries to attack other drones or the GS. So this adversary knows its own long term identity $ID_m$, sealed key $sk_m$, stable secret $s_m$, current counter $CC_m$, and pseudo identity $PID^{(t)}_m$. Also, it knows the shared group key $K_{grp}$, which is used for revocation broadcasts. However, it does not know the GS private signing key $sk_{GS}$, any other drone stable secret, ephemeral Kyber secrets, and $K_{sess}$ of sessions to which it is not a party. }

In this model, no party stores raw PUF outputs, stable secrets $s_i$, or long-term private keys in plaintext. All messages involving PUF responses are either obfuscated (DCS) or encrypted (AEAD). 
{Consequently, $D_m$ can authenticate as itself, but cannot forge a valid $Proof_j$ for another drone $D_j$, since this requires both $s_j$ and that session's $K_{sess}$. Also, $D_m$ cannot forge GS revocation broadcasts without $sk_{GS}$ and gains no advantage by knowing $K_{grp}$ since it does not provide authentication which comes from the GS signature $\sigma_{GS}$.}

\section{Performance Evaluation and Comparative Analysis}\label{sec:performance_compare}
In this section, we evaluate the performance of the proposed scheme by analyzing its communication, computation, and storage costs during the mutual authentication phases. We then provide a comparative analysis against representative state-of-the-art UAV authentication protocols. 


\subsection{Communication Cost}

Communication cost is measured as the total number of bits exchanged during the AKA phase of the protocol. We adopt the following standard bit-length assignments: SHA-256 hash output ($\mathit{PID}$) $= 256$ bits; Kyber-512 ephemeral public key $= 6400$ bits (800 bytes); Kyber-512 ciphertext $= 6144$ bits (768 bytes); scrambled challenge $C'_i = 128$ bits; nonce $= 128$ bits; AEAD-protected acknowledgment $= 384$ bits (256-bit ciphertext $+$ 128-bit authentication tag). 

{The above bit-length assignments follow the NIST Kyber KEM-512 specification for Kyber sizes \cite{alagic2024module}, the NIST secure hash standard for the SHA-256 hash output ($\mathit{PID}$) \cite{pub2012secure}, and RFC 8439 for the AEAD authentication tag length \cite{nir2018rfc}. Nonce and scrambled challenge sizes are protocol design parameters defined in Sections \ref{sec:syn_freshness_model} and \ref{sec:dcs}.}

\subsubsection{Drone--GS Authentication and Key Agreement Phase}  \label{sec:comm_dgs}

The communication costs for the four-message exchange of the Drone--GS AKA phase are detailed in Table~\ref{tab:comm_dgs}.

\begin{table}[htbp]
\centering
\caption{Communication Cost Analysis -- Drone--GS AKA Phase}
\label{tab:comm_dgs}
\setlength{\tabcolsep}{4pt}
\begin{tabular}{llcr}
\hline
\textbf{Flow} & \textbf{Messages Exchanged} & \textbf{Composition (bits)} & \textbf{Total (bits)} \\
\hline
$D_i \rightarrow \mathit{GS}$ & $\{\mathit{PID}^{(t)}_i,\; \mathit{pk}^{ep}_i\}$ & $256 + 6400$ & 6656 \\
$\mathit{GS} \rightarrow D_i$ & $\{c_{\mathit{GS}},\; N_{\mathit{GS}}\}$           & $6144 + 128$  & 6272 \\
$D_i \rightarrow \mathit{GS}$ & $\{C'_i,\; \mathit{Proof}_i\}$                     & $128 + 256$   & 384  \\
$\mathit{GS} \rightarrow D_i$ & \{AEAD acknowledgment\}                             & $256 + 128$   & 384  \\
\hline
\multicolumn{3}{r}{\textbf{Total communication cost (Drone--GS AKA phase)}} & \textbf{13696} \\
\hline
\end{tabular}
\end{table}

\subsubsection{D2D Mutual Authentication and Key Agreement Phase} \label{sec:comm_d2d_keyex}

The communication costs for the four-message exchange of the D2D AKA phase are detailed in Table~\ref{tab:comm_d2d}.

\begin{table*}[htbp]
\centering
\caption{Communication Cost Analysis -- D2D AKA Phase}
\label{tab:comm_d2d}
\setlength{\tabcolsep}{4pt}
\begin{tabular}{llcr}
\hline
\textbf{Flow} & \textbf{Messages Exchanged} & \textbf{Composition (bits)} & \textbf{Total (bits)} \\
\hline
$D_i \rightarrow D_j$ & $\{\mathit{PID}^{(t)}_i,\; \mathit{pk}^{ep}_i\}$ & $256 + 6400$ & 6656 \\
$D_j \rightarrow D_i$ & $\{\mathit{PID}^{(t)}_j,\; c_j,\; C'_j,\; N_j,\; \mathit{Proof}_j\}$ & $256+6144+128+128+256$ & 6912 \\
$D_i \rightarrow D_j$ & $\{H_i,\; \mathit{Proof}_i\}$ & $256 + 256$ & 512 \\
$D_j \rightarrow D_i$ & $\{H_j,\; \mathit{Proof}'_j\}$ & $256 + 256$ & 512 \\
\hline
\multicolumn{3}{r}{\textbf{Total communication cost (D2D AKA phase)}} & \textbf{14592} \\
\hline
\end{tabular}
\end{table*}

The dominant contributor to the communication overhead in both phases is the Kyber-512 ephemeral public key (800 bytes) and ciphertext (768 bytes), which together constitute 12{,}544 of the 13{,}696 bits in the Drone--GS phase. This overhead is structurally required by post-quantum key establishment and is the direct cost of achieving quantum-resistant session key agreement. Notably, Kyber-512 is the most bandwidth-efficient NIST-standardized post-quantum KEM; schemes based on BIKE or HQC would incur substantially greater overhead \cite{tawfeeq2025lightweight}. Once the Kyber handshake completes, the remaining protocol flows (scrambled challenge, PUF proof, AEAD acknowledgment) are relatively compact, contributing only 768 bits across three messages.

\subsection{Computation Cost}

The computation cost is measured as the total CPU time consumed during the online mutual authentication phase on resource-constrained UAV hardware. The Kyber-512 timings in Table~\ref{tab:op_timing} are converted from the pqm4 ARM Cortex-M4 cycle counts~\cite{kannwischer2019pqm4} at 168\,MHz. The hash and fuzzy extractor times are adopted from~\cite{algarni2024pslaps}, while the HKDF, AEAD, and PUF times are estimated values. The ephemeral Kyber key pair is generated offline and is therefore excluded from the online cost.

\begin{table}[htbp]
\centering
\caption{Operation Timing Values Used in the Cost Analysis}
\label{tab:op_timing}
\begin{tabular}{llc}
\hline
\textbf{Operation} & \textbf{Symbol} & \textbf{Time (ms)} \\
\hline
SHA-256 hash                     & $T_{H}$              & 0.0046  \\
Kyber-512 encapsulation          & $T_{\mathit{Kyber.Encap}}$  & 2.3270  \\
Kyber-512 decapsulation          & $T_{\mathit{Kyber.Decap}}$  & 2.5490  \\
HKDF key derivation              & $T_{\mathit{HKDF}}$         & 0.0090  \\
AEAD (ChaCha20-Poly1305)         & $T_{\mathit{AEAD}}$         & 0.1200  \\
PUF evaluation                   & $T_{\mathit{PUF}}$          & 0.1500  \\
Fuzzy extractor reproduction     & $T_{\mathit{FE.Rep}}$       & 3.8500  \\
XOR/ nonce generation           & $T_{\mathit{XOR}}$          & $\approx 0$ \\
\hline
\end{tabular}
\end{table}

\subsubsection{Drone--GS AKA Phase (Online Cost)}

The computation costs per entity during the Drone--GS AKA phase are shown in Table~\ref{tab:comp_dgs}.

\begin{table}[htbp]
\centering
\caption{Computation Cost Analysis -- Drone--GS AKA Phase}
\label{tab:comp_dgs}
\setlength{\tabcolsep}{4pt}
\begin{tabular}{lp{5.3cm}r}
\hline
\textbf{Participant} & \textbf{Operations} & \textbf{Cost (ms)} \\
\hline
GS &
$T_{\mathit{Kyber.Encap}} + 2T_{\mathit{HKDF}} + 3T_{H} + T_{\mathit{XOR}} + T_{\mathit{AEAD}}$ &
2.4788 \\[2pt]
Drone $D_i$ &
$T_{\mathit{Kyber.Decap}} + 2T_{\mathit{HKDF}} + T_{\mathit{PUF}} + T_{\mathit{FE.Rep}} + 2T_{H} + T_{\mathit{XOR}} + T_{\mathit{AEAD}}$ &
6.6962 \\
\hline
\multicolumn{2}{r}{\textbf{Total computation cost (Drone--GS AKA phase)}} & \textbf{9.1750} \\
\hline
\end{tabular}
\end{table}

The GS computation cost is dominated by Kyber-512 encapsulation ($2.327$ ms), while the drone cost is dominated by fuzzy extractor reproduction ($3.85$ ms) and Kyber-512 decapsulation ($2.549$ ms). PUF evaluation and hash operations together contribute less than $0.16$ ms. This asymmetry between GS and drone costs is acceptable in practice. The GS is usually considered to be a resource-rich infrastructure node, while the drone bears the higher but still bounded cost of PUF-based identity reconstruction.

\subsubsection{D2D AKA Phase (Online Cost)}

The computation costs for the Drone--Drone mutual authentication phase are shown in Table~\ref{tab:comp_d2d}.

\begin{table*}[htbp]
\centering
\caption{Computation Cost Analysis -- D2D AKA Phase}
\label{tab:comp_d2d}
\setlength{\tabcolsep}{4pt}
\begin{tabular}{lp{7.3cm}r}
\hline
\textbf{Participant} & \textbf{Operations} & \textbf{Cost (ms)} \\
\hline
Drone $D_i$ (initiator) &
$T_{\mathit{Kyber.Decap}} + T_{\mathit{HKDF}} + T_{\mathit{PUF}} + T_{\mathit{FE.Rep}} + 3T_{H} + T_{\mathit{XOR}}$ &
6.5718 \\[2pt]
Drone $D_j$ (responder) &
$T_{\mathit{Kyber.Encap}} + T_{\mathit{HKDF}} + T_{\mathit{PUF}} + T_{\mathit{FE.Rep}} + 3T_{H} + T_{\mathit{XOR}}$ &
6.3498 \\
\hline
\multicolumn{2}{r}{\textbf{Total computation cost (D2D AKA phase)}} & \textbf{12.9216} \\
\hline
\end{tabular}
\end{table*}

The near-symmetry in D2D costs shows that neither drone bears a disproportionate computational burden, which is critical for homogeneous swarm deployments where all nodes operate on identical hardware.

\subsection{Storage Cost}
 
Storage cost is measured as the total amount of persistent data
that each entity must retain across sessions to support the
authentication protocol.  Table~\ref{tab:storage} lists the
storage requirements of the proposed scheme for both the drone and
the GS.  The bit-length assumptions are: {Kyber-512 public key =
6400\,bits (800\,bytes) \cite{alagic2024module}, ML-DSA-44 public key = 10496\,bits (1312\,bytes) \cite{national2024module}, SHA-256 hash/commitment = 256\,bits \cite{pub2012secure},
sealed private key blob (AEAD ciphertext + authentication tag,
same size as underlying key) $\approx$ 512\,bits \cite{nir2018rfc} (Section \ref{SSC:PUF:Based:Key:Sealing}), fuzzy extractor
helper data $w_i$ = 256\,bits \cite{dodis2004fuzzy} (Section \ref{sec:PUF}), monotonic counter $CC_i$ = 32\,bits (Section \ref{sec:syn_freshness_model}),
group key $K_\mathrm{grp}$ = 256\,bits (Section \ref{sec:public_secret_key_mgmt}), enrollment challenge $Ch_i$
= 128\,bits (Section \ref{sec:network_model}), replay-cache entry (hashed tuple) = 256\,bits \cite{pub2012secure}, with a
bounded cache of 16 entries per entity (Section \ref{sec:syn_freshness_model}).}

{The fuzzy extractor helper data, monotonic counter, group key, enrollment challenge, and replay-cache bound are protocol design parameters, rather than externally mandated values.}
 
\begin{table}[!t]
\centering
\caption{Storage Cost Per Entity}
\label{tab:storage}
\begin{tabular}{l l r}
\hline
\textbf{Entity} & \textbf{Item} & \textbf{Bits} \\
\hline
\multirow{6}{*}{\textbf{Drone } $D_i$}
  & Enrollment challenge $Ch_i$        & 128 \\
  & Helper data $w_i$                  & 256 \\
  & Sealed private key $\mathit{Seal}_i$ & 512 \\
  & GS public key $pk_\mathrm{GS}$     & {10496} \\
  & Monotonic counter $CC_i$           & 32  \\
  & Group key $K_\mathrm{grp}$         & 256 \\
  & Replay cache (16 entries)          & 4096 \\
\cline{2-3}
  & \textbf{Drone total}               & \textbf{15776} \\
\hline
\multirow{5}{*}{\textbf{GS (per enrolled drone)}}
  & Identity $ID_i$                    & 256 \\
  & Long-term public key $pk_i$        & 6400 \\
  & Commitment $\mathit{Com}_i$        & 256 \\
  & Replay cache (16 entries / drone)  & 4096 \\
  & Group key $K_\mathrm{grp}$         & 256 \\
\cline{2-3}
  & \textbf{GS total (per drone)}      & \textbf{11264} \\
\hline
\end{tabular}
\end{table}
 
\textit{Drone storage:} {Each drone stores a total of $\approx$ 15776\,bits (1972\,bytes) of persistent data. The dominant cost is the GS long-term ML-DSA-44 public key $pk_{GS}$ (10496\,bits) \cite{national2024module}, which is required to verify revocation broadcasts.} All remaining per-drone data (challenge, helper data, sealed key, counter, group key, and replay cache) together amount to only 5280\,bits, confirming that the PUF
and DCS additions impose a negligible storage overhead over a plain Kyber-only scheme.
 
\textit{GS storage:}  The GS stores only the tuple
$(ID_i,\, pk_i,\, \mathit{Com}_i)$ per registered drone, resulting in a total of $\approx$11264\,bits per drone. Crucially, the GS never stores raw PUF responses, stable secrets $s_i$, or CRPs.  This is a major security advantage as even
a fully compromised GS leaks no information usable for drone impersonation or session-key reconstruction.

The proposed scheme's GS storage is bounded, predictable, and capture-resistant.  Its per-drone footprint of 11264\,bits is higher than that of simplified schemes such as Choi~\textit{et al.} \cite{choi2025puf} because it includes a full Kyber-512 long-term public key. This cost is however offset by the significantly stronger security guarantee set provided, in particular post-quantum resistance, which those schemes do not offer. In summary, the proposed scheme eliminates the unbounded CRP database required by conventional PUF schemes, replaces it with a fixed-size commitment of 256\,bits, and maintains a drone-side footprint small enough for typical embedded UAV storage (approximately 1972\,bytes, well within the flash capacity of Cortex-M4-class microcontrollers).

\subsection{Comparative Analysis}

\begin{figure}[htbp]
  \centering
  \includegraphics[width=1\linewidth]{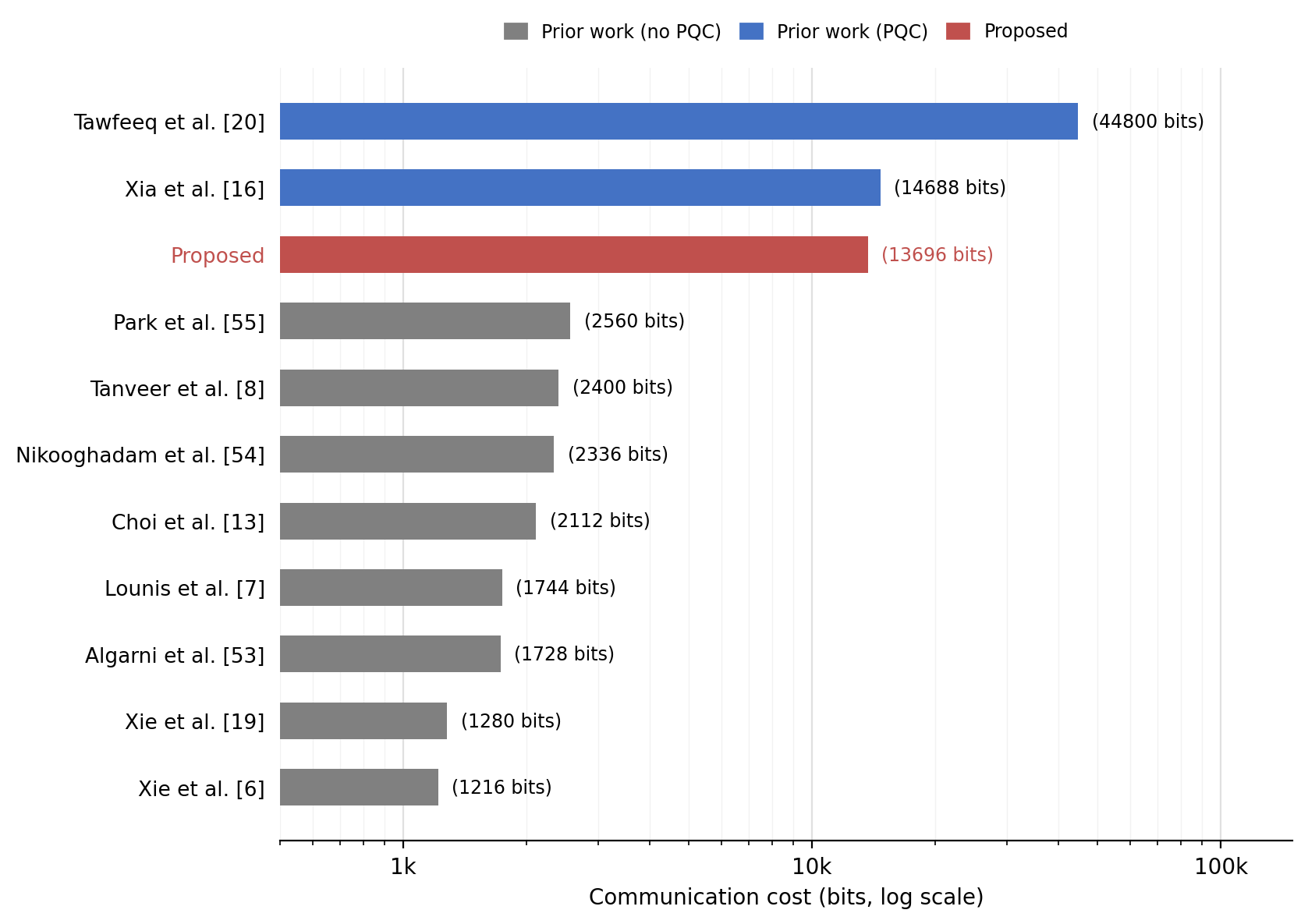}
  \caption{Comparative analysis of communication cost (bits) for the Drone–GS AKA phase.}
  \label{fig:comparison}
\end{figure}

\begin{figure}[htbp]
  \centering
  \includegraphics[width=1\linewidth]{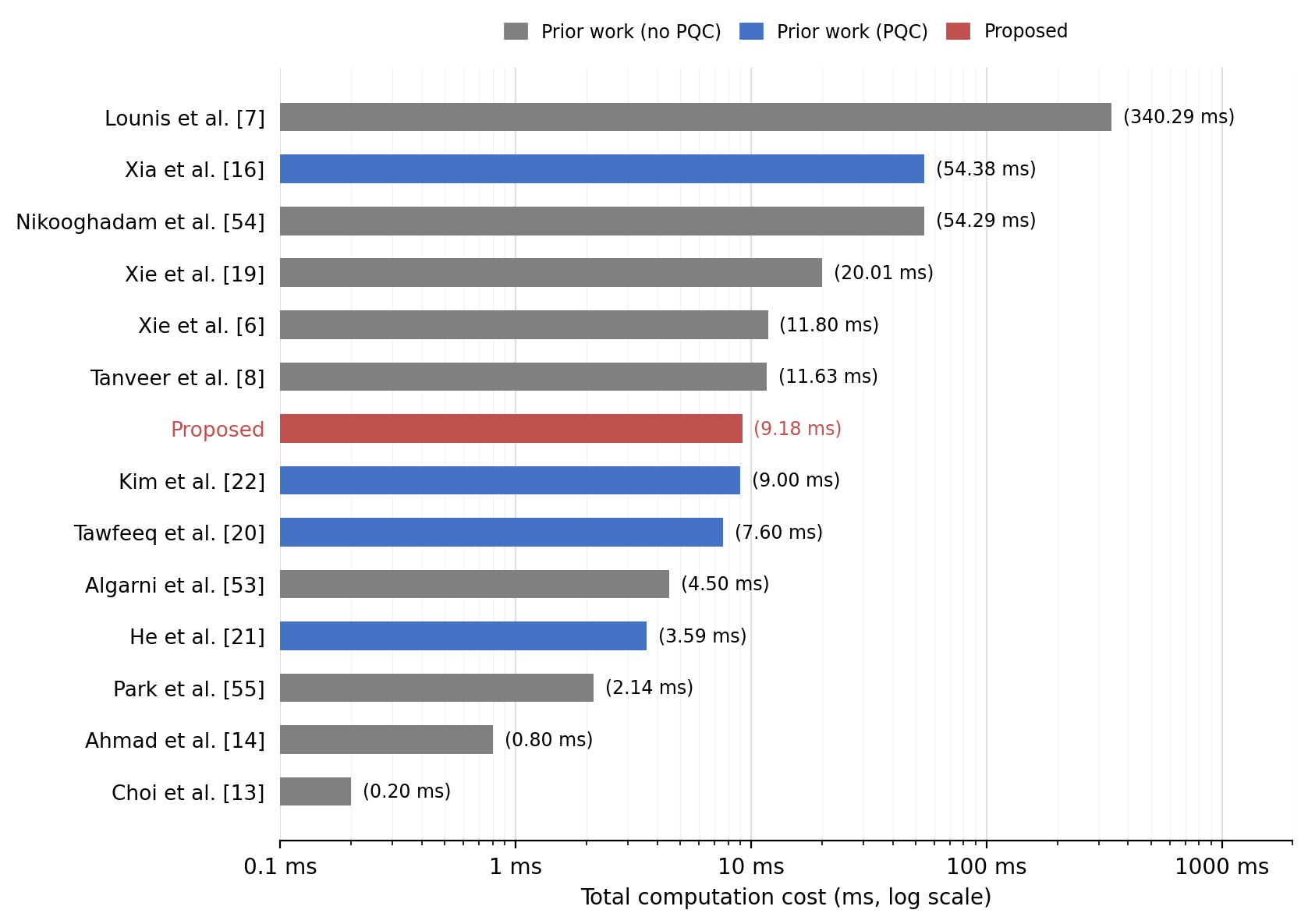}
  \caption{Comparative analysis of computation cost (ms) for the Drone–GS AKA phase.}
  \label{fig:comparison2}
\end{figure}

Table \ref{tab:security_comparison} compares the security aspects of the proposed scheme against thirteen representative IoD authentication protocols, spanning classical hash/XOR designs, PUF-based schemes, and post-quantum approaches. The communication and computation costs are compared in Fig.~\ref{fig:comparison} and Fig.~\ref{fig:comparison2} and GS storage per enrolled drone in Table \ref{tab:storage_cmp}.

\begin{table*}[!t]
\centering
\caption{Security Comparison With State-of-the-Art Schemes
($\checkmark$ = Supported; $\times$ = Not Supported)}
\label{tab:security_comparison}
\resizebox{\textwidth}{!}{%
\begin{tabular}{l c c c c c c c}
\hline
\textbf{Scheme} &
\textbf{\shortstack{Post-\\quantum}} &
\textbf{PUF} &
\textbf{\shortstack{Forward\\secrecy}} &
\textbf{Anonymity} &
\textbf{\shortstack{D2D\\auth.}} &
\textbf{\shortstack{Anti-\\modeling}} &
\textbf{\shortstack{Mutual\\auth.}} \\
\hline
Xie et al.~\cite{xie2025lightweight}         & $\times$ & $\checkmark$ & $\times$ & $\times$ & $\checkmark$ & $\times$ & $\checkmark$ \\
Lounis et al.~\cite{lounis2022d2d}      & $\times$ & $\checkmark$ & $\times$ & $\times$ & $\checkmark$ & $\times$ & $\checkmark$ \\
Tanveer et al.~\cite{tanveer2024paf}    & $\times$ & $\checkmark$ & $\checkmark$ & $\times$ & $\times$ & $\times$ & $\checkmark$ \\
Choi et al.~\cite{choi2025puf}        & $\times$ & $\checkmark$ & $\checkmark$ & $\times$ & $\times$ & $\times$ & $\checkmark$ \\
Ahmad et al.~\cite{ahmad2026secure}     & $\times$ & $\checkmark$ & $\times$ & $\times$ & $\checkmark$ & $\times$ & $\checkmark$ \\
Xia et al.~\cite{xia2024quantum}           & $\checkmark$ & $\times$ & $\checkmark$ & $\times$ & $\times$ & $\times$ & $\checkmark$ \\
Xie et al.~\cite{xie2025puf}          & $\times$ & $\checkmark$ & $\times$ & $\checkmark$ & $\times$ & $\times$ & $\checkmark$ \\
Tawfeeq et al.~\cite{tawfeeq2025lightweight}  & $\checkmark$ & $\times$ & $\times$ & $\times$ & $\times$ & $\times$ & $\checkmark$ \\
He et al.~\cite{he2025post}            & $\checkmark$ & $\times$ & $\checkmark$ & $\times$ & $\checkmark$ & $\times$ & $\checkmark$ \\
Kim et al.~\cite{kim2026pqrid}      & $\checkmark$ & $\times$ & $\times$ & $\checkmark$ & $\times$ & $\times$ & $\checkmark$ \\
Algarni et al.~\cite{algarni2024pslaps}     & $\times$ & $\times$ & $\checkmark$ & $\times$ & $\times$ & $\times$ & $\checkmark$ \\
Nikooghadam et al.~\cite{nikooghadam2021provably} & $\times$ & $\times$ & $\checkmark$ & $\times$ & $\times$ & $\times$ & $\checkmark$ \\
Park et al.~\cite{park2023provably}          & $\times$ & $\checkmark$ & $\checkmark$ & $\times$ & $\times$ & $\times$ & $\checkmark$ \\

\hline
\textbf{Proposed}                       & $\checkmark$ & $\checkmark$ & $\checkmark$ & $\checkmark$ & $\checkmark$ & $\checkmark$ & $\checkmark$ \\
\hline
\end{tabular}}
\end{table*}
\subsubsection{Computation Cost}

The proposed scheme achieves a total online computation cost of $9.1750$ ms (Drone--GS phase) as shown in Table \ref{tab:comp_dgs}. This represents an improvement of $97.30\%$ over \cite{lounis2022d2d} ($340.294$ ms), $83.13\%$ over \cite{xia2024quantum} ($54.379$ ms), 83.10\% over \cite{nikooghadam2021provably} (54.292\,ms), 54.15\% over \cite{xie2025puf} (20.01\,ms), 22.27\% over \cite{xie2025lightweight} (11.804\,ms), and 21.10\% over \cite{tanveer2024paf} (11.628\,ms).

{Conversely, \cite{choi2025puf}, \cite{ahmad2026secure}, \cite{tawfeeq2025lightweight}, \cite{he2025post}, \cite{kim2026pqrid}, \cite{algarni2024pslaps}, and \cite{park2023provably} achieve lower computational cost than the proposed scheme (Fig. \ref{fig:comparison2}). Among these seven schemes, \cite{choi2025puf}, \cite{ahmad2026secure},  \cite{algarni2024pslaps}, and \cite{park2023provably} omit post-quantum key establishment, one of the costliest operations in the proposed protocol, while \cite{tawfeeq2025lightweight}, \cite{he2025post}, and \cite{kim2026pqrid} are post-quantum but omit the PUF-based step, leaving them vulnerable to cloning and impersonation attacks. As summarized in Table \ref{tab:security_comparison}, none of these seven schemes simultaneously provides post-quantum resistance, PUF-based hardware trust, anonymity, D2D authentication, forward secrecy, and anti-modeling protection, all of which the proposed scheme achieves. Their lower computational cost therefore reflects a narrower security-feature set rather than a more efficient realization of the same guarantees.}

\subsubsection{Communication Cost}

{The proposed scheme incurs $13696$ bits of communication overhead in the Drone--GS AKA phase, as shown in Table \ref{tab:comm_dgs}. The scheme of \cite{xia2024quantum} requires a higher overhead of $14688$ bits, so the proposed scheme achieves a $6.75\%$ reduction in communication cost. The scheme of \cite{tawfeeq2025lightweight} also incurs a higher overhead of approximately $44800$ bits, owing to its CRYSTALS-Dilithium signatures, so the proposed scheme achieves a $69.43\%$ reduction. In addition to this, the proposed scheme provides a strictly broader security guarantee set than both \cite{xia2024quantum} and \cite{tawfeeq2025lightweight}, including PUF-based hardware trust, DCS-based anti-modeling protection, anonymity, and D2D authentication.}

{Conversely, the schemes of \cite{xie2025lightweight}, \cite{lounis2022d2d}, \cite{tanveer2024paf}, \cite{choi2025puf}, \cite{xie2025puf}, \cite{algarni2024pslaps}, \cite{nikooghadam2021provably}, and \cite{park2023provably} achieve lower communication cost than the proposed scheme (Fig. \ref{fig:comparison}). This is a direct consequence of their reliance on classical key establishment. None of them transmits a Kyber-$512$ public key or ciphertext, which together account for $12544$ of the $13696$ bits in the proposed scheme's Drone-GS phase (Section \ref{sec:comm_d2d_keyex}). Consequently, as indicated in Table \ref{tab:security_comparison}, all of them are broken by a sufficiently powerful quantum adversary, and lack anti-modeling protection against PUF-based machine learning attacks. The scheme of \cite{tawfeeq2025lightweight} is post-quantum, but it doesn't incorporate PUF-based authentication and omit the scrambled challenge and PUF proof exchange ($C'_i$, $Proof_i$) that binds the proposed protocol's sessions to physical hardware. In every case, the lower cost of these eight schemes is due to no post-quantum resistance, hardware rooted trust, and anti-modeling protection which the proposed scheme provides simultaneously.}

\subsubsection{Storage Cost}  
 Table~\ref{tab:storage_cmp} compares per-drone GS storage requirements across representative IoD schemes. Classical PUF-only schemes such as Tanveer~\textit{et al.} \cite{tanveer2024paf} and Lounis~\textit{et al.} \cite{lounis2022d2d} require the GS to store complete CRP databases to perform verification. For a 128-bit challenge with $N$ pairs, this amounts to $N \times 256$\,bits, which is a potentially unbounded cost that grows with usage and represents a severe vulnerability if the GS is captured.  Post-quantum signature-based schemes such as Tawfeeq~\textit{et al.} \cite{tawfeeq2025lightweight} must store full Dilithium public keys (1312\,bytes each for
Dilithium2), leading to a per-drone GS footprint of approximately 13696\,bits before replay-protection data which is higher than the proposed scheme. Schemes based solely on Kyber without PUFs, such as Xia~\textit{et al.} \cite{xia2024quantum}, must maintain a fresh ephemeral public key per session or, in long-term deployments, a long-term public key plus session metadata. {Similarly, He~\textit{et al.}~\cite{he2025post} omits PUF-based hardware binding and proposes a Kyber-based group key agreement scheme for drone swarms in which identity is managed through a SMT, requiring the GS to store a Kyber public key together with the SMT root data, yielding a per-drone footprint of approximately 7424 bits.} PUF-free schemes incidentally expose the GS to greater risk from compromise since public-key material alone suffices for impersonation if the corresponding private key is later extracted.  The proposed scheme achieves a compact, bounded, and physically compromise-resistant GS storage footprint by substituting the unbounded CRP database of conventional PUF schemes with a single 256-bit commitment $\mathit{Com}_i = H(ID_i\,\|\,pk_i\,\|\,s_i)$ per drone.  This commitment is one-way and therefore useless to an
attacker even if the GS is fully compromised.
 
\begin{table}[!t]
\centering
\caption{Comparative GS Storage Cost Per Enrolled Drone}
\label{tab:storage_cmp}
\begin{tabular}{l r l}
\hline
\textbf{Scheme} & \textbf{GS bits / drone} & \textbf{Note} \\
\hline
Lounis et al.~\cite{lounis2022d2d}               & $\gg 256N$      & Full CRP DB \\
Tanveer et al.~\cite{tanveer2024paf}               & $\gg 256N$      & Full CRP DB ($N$ pairs) \\
Choi et al.~\cite{choi2025puf} & $\approx$5120   & PUF commitment + pk \\
Xia et al.~\cite{xia2024quantum}            & $\approx$7168   & Kyber pk + identity hash \\
Tawfeeq et al.~\cite{tawfeeq2025lightweight} & $\approx$13696 & Dilithium2 pk + metadata \\
He et al.~\cite{he2025post}     & $\approx$7424   & Kyber-512 pk + SMT root hash \\
\textbf{Proposed}               & \textbf{11264}  & \textbf{Bounded, no CRP DB} \\
                                &                 & \textbf{GS capture-safe} \\
\hline
\end{tabular}
\end{table}

\subsubsection{Security Feature Coverage}

The proposed scheme is the only protocol in the comparison (Table \ref{tab:security_comparison}) that simultaneously satisfies all seven evaluated properties of post-quantum resistance, PUF-based hardware binding, forward secrecy, anonymity/unlinkability, D2D authentication without GS involvement, anti-modeling protection via DCS, and mutual authentication. No existing scheme in the comparison satisfies more than four of these properties simultaneously. This positions the proposed protocol as the most feature-complete solution for IoD environments operating under adversarial and quantum-capable conditions.


\section{Conclusions and Future Work}\label{sec:conclusion}

This paper presented a lightweight and quantum-resilient AKA framework for secure IoD environments. The proposed protocol integrates CRYSTALS-Kyber post-quantum cryptography, PUFs, fuzzy extractors, DCS, and PID mechanisms to provide secure and anonymous UAV communication. Unlike conventional UAV authentication schemes, the proposed framework simultaneously achieves post-quantum resistance, hardware-rooted trust, forward secrecy, anonymity preservation, and resistance against replay, impersonation, and physical capture attacks.

The integration of DCS with Kyber-derived session material ensures that exposed challenge--response observations remain unusable for machine-learning-based PUF modeling attacks. Furthermore, the protocol eliminates the need for storing raw CRP databases or plaintext long-term secrets, significantly reducing the impact of drone or GS compromise. Formal security analysis under the RoR model demonstrated the semantic security of the established session keys under MLWE hardness and PUF unpredictability assumptions.

Performance evaluation showed that the proposed scheme remains computationally practical for resource-constrained UAV hardware, while providing stronger security guarantees than existing approaches. Hence, the proposed protocol offers a practical  authentication solution for next-generation UAV networks operating in adversarial and quantum-capable environments.

For future investigation, since the proposed scheme uses Kyber-512, future work could evaluate the trade-offs of upgrading to Kyber-768 or Kyber-1024 for higher security levels in mission-critical deployments. Also, the proposed framework can also be applied to related constrained-device domains such as the Internet of Vehicles (IoV) and satellite-based communication networks, providing a natural generalization.

\bibliographystyle{IEEEtran} \bibliography{ref}

@article{xie2025lightweight,
  title={Lightweight drone-to-ground station and drone-to-drone authentication scheme for {Internet of Drones}},
  author={Xie, Qi and Zhang, Jilin},
  journal={Symmetry},
  volume={17},
  number={4},
  pages={556},
  year={2025},
  publisher={MDPI}
}

@article{lounis2022d2d,
  title={{D2D-MAP}: A drone to drone authentication protocol using physical unclonable functions},
  author={Lounis, Karim and Ding, Steven HH and Zulkernine, Mohammad},
  journal={IEEE Transactions on Vehicular Technology},
  volume={72},
  number={4},
  pages={5079--5093},
  year={2022},
  publisher={IEEE}
}

@article{tanveer2024paf,
  title={{PAF-IoD: {PUF}-enabled authentication framework for the Internet of Drones}},
  author={Tanveer, Muhammad and Aldosary, Abdallah and Das, Ashok Kumar and Aldossari, Saud Alhajaj and Chaudhry, Shehzad Ashraf and others},
  journal={IEEE Transactions on Vehicular Technology},
  volume={73},
  number={7},
  pages={9560--9574},
  year={2024},
  publisher={IEEE}
}

@article{xia2024quantum,
  title={A quantum-resistant identity authentication and key agreement scheme for {UAV} networks based on {Kyber} algorithm},
  author={Xia, Tao and Wang, Menglin and He, Jun and Yang, Gang and Fan, Linna and Wei, Guoheng},
  journal={Drones},
  volume={8},
  number={8},
  pages={359},
  year={2024},
  publisher={MDPI}
}

@article{mishra2023quantum,
  title={Quantum-safe secure and authorized communication protocol for {Internet of Drones}},
  author={Mishra, Dheerendra and Singh, Mrityunjay and Rewal, Purva and Pursharthi, Komal and Kumar, Neeraj and Barnawi, Ahmed and Rathore, Rajkumar Singh},
  journal={IEEE Transactions on Vehicular Technology},
  volume={72},
  number={12},
  pages={16499--16507},
  year={2023},
  publisher={IEEE}
}

@inproceedings{aissaoui2023authenticating,
  title={Authenticating civil {UAV} communications with post-quantum digital signatures},
  author={Aissaoui, Ridwane and Deneuville, Jean-Christophe and Guerber, Christophe and Pirovano, Alain},
  booktitle={2023 IEEE/AIAA 42nd Digital Avionics Systems Conference (DASC)},
  pages={1--9},
  year={2023},
  organization={IEEE}
}

@article{han2024enhanced,
  title={An enhanced multifactor authentication and key agreement protocol in industrial {Internet of Things}},
  author={Han, Yiran and Guo, Hua and Liu, Jianwei and Ehui, Brou Bernard and Wu, Yapeng and Li, Sijia},
  journal={IEEE {Internet of Things} Journal},
  volume={11},
  number={9},
  pages={16243--16254},
  year={2024},
  publisher={IEEE}
}

@article{khan2024future,
  title={Future-proofing security for {UAVs} with post-quantum cryptography: A review},
  author={Khan, Muhammad Asghar and Javaid, Shumaila and Mohsan, Syed Agha Hassnain and Tanveer, Muhammad and Ullah, Insaf},
  journal={IEEE Open Journal of the Communications Society},
  volume={5},
  pages={6849--6871},
  year={2024},
  publisher={IEEE}
}

@article{hussain2023user,
  author={Hussain, Sajid and Farooq, Muhammad and Alzahrani, Bander A. and Albeshri, Aiiad and Alsubhi, Khalid and Chaudhry, Shehzad Ashraf},
  journal={IEEE Access}, 
  title={An Efficient and Reliable User Access Protocol for {Internet of Drones}}, 
  year={2023},
  volume={11},
  number={},
  pages={59688--59700},
  doi={10.1109/ACCESS.2023.3284832}}

@article{garcia2023tesla,
  author={García, Julio César Pérez and Benslimane, Abderrahim and Braeken, An and Su, Zhou},
  journal={IEEE {Internet of Things} Journal}, 
  title={{$\mu$TESLA}-Based Authentication for Reliable and Secure Broadcast Communications in {IoD} Using Blockchain}, 
  year={2023},
  volume={10},
  number={20},
  pages={18400--18413},
  doi={10.1109/JIOT.2023.3280124}
  }

@article{cho2020sentinel,
  title={{SENTINEL}: A secure and efficient authentication framework for unmanned aerial vehicles},
  author={Cho, Geumhwan and Cho, Junsung and Hyun, Sangwon and Kim, Hyoungshick},
  journal={Applied Sciences},
  volume={10},
  number={9},
  pages={3149},
  year={2020},
  publisher={MDPI}
}

@article{yu2023laka,
  title={{LAKA-UAV}: Lightweight authentication and key agreement scheme for cloud-assisted unmanned aerial vehicle using blockchain in flying ad-hoc networks},
  author={Yu, Sungjin and Lee, Joonyoung and Sutrala, Anil Kumar and Das, Ashok Kumar and Park, Youngho},
  journal={Computer Networks},
  volume={224},
  pages={109612},
  year={2023},
  publisher={Elsevier}
}

@article{bhattarai2024IoD,
  author={Bhattarai, Image and Pu, Cong and Choo, Kim-Kwang Raymond and Korać, Dragan},
  journal={IEEE {Internet of Things} Journal}, 
  title={A Lightweight and Anonymous Application-Aware Authentication and Key Agreement Protocol for the {Internet of Drones}}, 
  year={2024},
  volume={11},
  number={11},
  pages={19790--19803},
  doi={10.1109/JIOT.2024.3367799}}

@article{pan2024privacy,
  title={Privacy-preserving location authentication for low-altitude {UAVs}: A blockchain-based approach},
  author={Pan, Hengchang and Wang, Yuanshuo and Wang, Wei and Cao, Ping and Ye, Fangwei and Wu, Qihui},
  journal={Security and Safety},
  volume={3},
  pages={2024004},
  year={2024},
  publisher={EDP Sciences and CSPM}
}

@article{ibrahim2024resilience,
  title={Resilience evaluation of memristor based {PUF} against machine learning attacks},
  author={Ibrahim, Hebatallah M and Skovorodnikov, Heorhii and Alkhzaimi, Hoda},
  journal={Scientific Reports},
  volume={14},
  number={1},
  pages={23962},
  year={2024},
  publisher={Nature Publishing Group UK London}
}

@inproceedings{mefgouda2024lPUF,
  title={{LPUF-AuthNet}: A lightweight {PUF}-based {IoT} authentication via tandem neural networks and split learning},
  author={Mefgouda, Brahim and Khan, Raviha and Alhussein, Omar and Saleh, Hani and Eldeeb, Hossien B and Pandey, Anshul and Muhaidat, Sami},
  booktitle={2024 IEEE Globecom Workshops (GC Wkshps)},
  pages={1--7},
  year={2024},
  organization={IEEE}
}

@inproceedings{gupta2022secure,
  title={Secure, anonymity-preserving and lightweight mutual authentication and key agreement protocol for home automation {IoT} networks},
  author={Gupta, Akash and Kasbekar, Gaurav S},
  booktitle={2022 14th International Conference on COMmunication Systems \& NETworkS (COMSNETS)},
  pages={375--383},
  year={2022},
  organization={IEEE}
}

@article{dolev1983security,
  title={On the security of public key protocols},
  author={Dolev, Danny and Yao, Andrew},
  journal={IEEE Transactions on Information Theory},
  volume={29},
  number={2},
  pages={198--208},
  year={1983},
  publisher={IEEE}
}

@article{ebrahimabadi2024robust,
  title={Robust and Lightweight Challenge Obfuscation Mechanism for Anti-modeling Protection of Arbiter-{PUFs}},
  author={Ebrahimabadi, Mohammad and Younis, Mohamed and Mehjabin, Suhee Sanjana and Tekeoglu, Ali and Sookoor, Tamim I and Karimi, Naghmeh},
  journal={Journal of Hardware and Systems Security},
  volume={8},
  number={4},
  pages={205--216},
  year={2024},
  publisher={Springer}
}

@article{10.1145/2818186,
author = {Delvaux, Jeroen and Peeters, Roel and Gu, Dawu and Verbauwhede, Ingrid},
title = {A Survey on Lightweight Entity Authentication with Strong {PUFs}},
year = {2015},
issue_date = {November 2015},
publisher = {Association for Computing Machinery},
address = {New York, NY, USA},
volume = {48},
number = {2},
issn = {0360-0300},
doi = {10.1145/2818186},
journal = {ACM Computing Surveys},
month = oct,
articleno = {26},
numpages = {42}
}

@article{algarni2024pslaps,
  title={{PSLAPS-IoD}: A provable secure and lightweight authentication protocol for securing {Internet-of-Drones} ({IoD}) environment},
  author={Algarni, Fahad and Jan, Saeed Ullah},
  journal={IEEE Access},
  volume={12},
  pages={45948--45960},
  year={2024},
  publisher={IEEE}
}

@article{nikooghadam2021provably,
  title={A provably secure and lightweight authentication scheme for {Internet of Drones} for smart city surveillance},
  author={Nikooghadam, Mahdi and Amintoosi, Haleh and Islam, SK Hafizul and Moghadam, Mostafa Farhadi},
  journal={Journal of Systems Architecture},
  volume={115},
  pages={101955},
  year={2021},
  publisher={Elsevier}
}

@article{park2023provably,
  title={Provably secure mutual authentication and key agreement scheme using {PUF} in {Internet of Drones} deployments},
  author={Park, Yohan and Ryu, Daeun and Kwon, Deokkyu and Park, Youngho},
  journal={Sensors},
  volume={23},
  number={4},
  pages={2034},
  year={2023},
  publisher={MDPI}
}

@inproceedings{dodis2004fuzzy,
  title={Fuzzy extractors: How to generate strong keys from biometrics and other noisy data},
  author={Dodis, Yevgeniy and Reyzin, Leonid and Smith, Adam},
  booktitle={International Conference on the Theory and Applications of Cryptographic Techniques},
  pages={523--540},
  year={2004},
  organization={Springer}
}

@misc{langlois2012modulelattices,
      author = {Adeline Langlois and Damien Stehle},
      title = {Worst-Case to Average-Case Reductions for Module Lattices},
      howpublished = {Cryptology {ePrint} Archive, Paper 2012/090},
      year = {2012},
      url = {https://eprint.iacr.org/2012/090}
}

@article{blanchet2016modeling,
author = {Blanchet, Bruno},
title = {Modeling and verifying security protocols with the applied pi calculus and {ProVerif}},
year = {2016},
issue_date = {31 10 2016},
publisher = {Now Publishers Inc.},
address = {Hanover, MA, USA},
volume = {1},
number = {1--2},
issn = {2474-1558},
doi = {10.1561/3300000004},
journal = {Foundations and Trends in Privacy and Security},
pages = {1--135},
numpages = {135}
}

@article{abadi2001mobile,
  title={Mobile values, new names, and secure communication},
  author={Abadi, Mart{\'\i}n and Fournet, C{\'e}dric},
  journal={ACM SIGPLAN Notices},
  volume={36},
  number={3},
  pages={104--115},
  year={2001},
  publisher={ACM New York, NY, USA}
}

@article{xie2025PUF,
  title={{PUF}-based secure and efficient anonymous authentication protocol for {UAV} towards cross-domain environments},
  author={Xie, Qi and Wang, Haohua},
  journal={Drones},
  volume={9},
  number={4},
  pages={260},
  year={2025},
  publisher={MDPI}
}

@article{tawfeeq2025lightweight,
  title={Lightweight and quantum-resistant authentication for the {Internet of Drones} {(IoD) using {Dilithium}} signatures},
  author={Tawfeeq, Nawar Hayder and Yousif, Mohammed and Al-Shareeda, Mahmood A and Almaiah, Mohammed Amin and Shehab, Rami},
  journal={International Journal of Innovative Research and Scientific Studies},
  volume={8},
  number={2},
  pages={2842--2853},
  year={2025}
}

@article{he2025post,
  title={A Post-Quantum Authentication and Key Agreement Scheme for Drone Swarms},
  author={He, Linlin and Zhao, Meng and Wang, Xu’an and Wang, Jue and Wang, Zhenyu and Liu, Shuanggen},
  journal={Electronics},
  volume={14},
  number={17},
  pages={3364},
  year={2025},
  publisher={MDPI}
}

@article{choi2025PUF,
  title={A {PUF}-based secure authentication and key agreement scheme for the {Internet of Drones}},
  author={Choi, Jihye and Son, Seunghwan and Kwon, Deokkyu and Park, Youngho},
  journal={Sensors},
  volume={25},
  number={3},
  pages={982},
  year={2025},
  publisher={MDPI}
}

@article{kim2026pqrid,
  title={{pqRID}: Compact, Anonymous Remote Identification and Post-Quantum Authentication for Drones},
  author={Kim, Jane and Lee, YeoNyeong and Kang, Jung-Hun and Seo, Seung-Hyun},
  journal={IEEE {Internet of Things} Journal},
  year={2026},
  publisher={IEEE}
}

@article{ahmad2026secure,
  title={Secure and Decentralised Swarm Authentication Using Hardware Security Primitives},
  author={Ahmad, Sagir Muhammad and Honarvar Shakibaei Asli, Barmak},
  journal={Electronics},
  volume={15},
  number={2},
  pages={423},
  year={2026},
  publisher={MDPI}
}

@inproceedings{shor1994algorithms,
  title={Algorithms for quantum computation: discrete logarithms and factoring},
  author={Shor, Peter W},
  booktitle={Proceedings of the 35th Annual Symposium on Foundations of Computer Science},
  pages={124--134},
  year={1994},
  organization={IEEE}
}

@techreport{alagic2024module,
  title={{Module-Lattice-Based Key-Encapsulation Mechanism Standard}},
  author={{National Institute of Standards and Technology}},
  institution={U.S. Department of Commerce},
  type={Federal Information Processing Standards Publication},
  number={FIPS 203},
  year={2024},
  month=aug,
  doi={10.6028/NIST.FIPS.203}
}

@article{pappu2002physical,
  title={Physical one-way functions},
  author={Pappu, Ravikanth and Recht, Ben and Taylor, Jason and Gershenfeld, Neil},
  journal={Science},
  volume={297},
  number={5589},
  pages={2026--2030},
  year={2002},
  publisher={American Association for the Advancement of Science}
}

@inproceedings{gassend2002silicon,
  title={Silicon physical random functions},
  author={Gassend, Blaise and Clarke, Dwaine and Van Dijk, Marten and Devadas, Srinivas},
  booktitle={{Proceedings of the 9th ACM Conference on Computer and Communications Security}},
  pages={148--160},
  year={2002}
}

@inproceedings{ruhrmair2010modeling,
  title={Modeling attacks on physical unclonable functions},
  author={R{\"u}hrmair, Ulrich and Sehnke, Frank and S{\"o}lter, Jan and Dror, Gideon and Devadas, Srinivas and Schmidhuber, J{\"u}rgen},
  booktitle={P{roceedings of the 17th ACM conference on Computer and Communications Security}},
  pages={237--249},
  year={2010}
}

@article{mozaffari2019tutorial,
  title={A tutorial on {UAVs} for wireless networks: Applications, challenges, and open problems},
  author={Mozaffari, Mohammad and Saad, Walid and Bennis, Mehdi and Nam, Young-Han and Debbah, M{\'e}rouane},
  journal={IEEE Communications Surveys \& Tutorials},
  volume={21},
  number={3},
  pages={2334--2360},
  year={2019},
  publisher={IEEE}
}

@article{shakhatreh2019unmanned,
  title={Unmanned aerial vehicles ({UAVs}): A survey on civil applications and key research challenges},
  author={Shakhatreh, Hazim and Sawalmeh, Ahmad H and Al-Fuqaha, Ala and Dou, Zuochao and Almaita, Eyad and Khalil, Issa and Othman, Noor Shamsiah and Khreishah, Abdallah and Guizani, Mohsen},
  journal={IEEE Access},
  volume={7},
  pages={48572--48634},
  year={2019},
  publisher={IEEE}
}

@inproceedings{bellare1993entity,
  title={Entity authentication and key distribution},
  author={Bellare, Mihir and Rogaway, Phillip},
  booktitle={Annual International Cryptology Conference},
  pages={232--249},
  year={1993},
  organization={Springer}
}

@article{mall2022PUF,
  title={{{PUF}-based authentication and key agreement protocols for {IoT}, WSNs, and smart grids: A comprehensive survey}},
  author={Mall, Priyanka and Amin, Ruhul and Das, Ashok Kumar and Leung, Mark T and Choo, Kim-Kwang Raymond},
  journal={IEEE {Internet of Things} Journal},
  volume={9},
  number={11},
  pages={8205--8228},
  year={2022},
  publisher={IEEE}
}

@article{nguyen2018interpose,
  title={{The interpose {PUF}: Secure {PUF} design against state-of-the-art machine learning attacks}},
  author={Nguyen, Phuong Ha and Sahoo, Durga Prasad and Jin, Chenglu and Mahmood, Kaleel and R{\"u}hrmair, Ulrich and Van Dijk, Marten},
  journal={Cryptology ePrint Archive},
  year={2018}
}

@inproceedings{suh2007physical,
  title={Physical unclonable functions for device authentication and secret key generation},
  author={Suh, G Edward and Devadas, Srinivas},
  booktitle={Proceedings of the 44th Annual Design Automation Conference},
  pages={9--14},
  year={2007}
}

@inproceedings{cremers2023automated,
  title={Automated Analysis of Protocols that use Authenticated Encryption: How Subtle {AEAD} Differences can impact Protocol Security},
  author={Cremers, Cas and Dax, Alexander and Jacomme, Charlie and Zhao, Mang},
  booktitle={32nd USENIX Security Symposium (USENIX Security 23)},
  pages={5935--5952},
  year={2023}
}

@techreport{pub2012secure,
  title={{Secure Hash Standard (SHS)}},
  author={{National Institute of Standards and Technology}},
  institution={U.S. Department of Commerce},
  type={Federal Information Processing Standards Publication},
  number={FIPS 180-4},
  year={2015},
  doi={10.6028/NIST.FIPS.180-4}
}

@misc{nir2018rfc,
  title={{RFC 8439: ChaCha20 and Poly1305 for {IETF} protocols}},
  author={Nir, Yoav and Langley, Adam},
  year={2018},
  publisher={RFC Editor},
  doi={10.17487/RFC8439}
}

@techreport{national2024module,
  title={{Module-Lattice-Based Digital Signature Standard}},
  author={{National Institute of Standards and Technology}},
  institution={U.S. Department of Commerce},
  type={Federal Information Processing Standards Publication},
  number={FIPS 204},
  year={2024},
  doi={10.6028/NIST.FIPS.204}
}

@book{katz2020introduction,
  author    = {Katz, Jonathan and Lindell, Yehuda},
  title     = {Introduction to Modern Cryptography},
  edition   = {3rd},
  publisher = {Chapman and Hall/CRC},
  address   = {Boca Raton, FL, USA},
  year      = {2020},
  isbn      = {978-0-8153-5436-9}
}

@inproceedings{bellare1993random,
  title={Random oracles are practical: A paradigm for designing efficient protocols},
  author={Bellare, Mihir and Rogaway, Phillip},
  booktitle={Proceedings of the 1st ACM Conference on Computer and Communications Security},
  pages={62--73},
  year={1993}
}

@article{alkatheiri2022lightweight,
  title={A lightweight authentication scheme for a network of unmanned aerial vehicles ({UAVs}) by using physical unclonable functions},
  author={Alkatheiri, Mohammed Saeed and Saleem, Sajid and Alqarni, Mohammed Ali and Aseeri, Ahmad O and Chauhdary, Sajjad Hussain and Zhuang, Yu},
  journal={Electronics},
  volume={11},
  number={18},
  pages={2921},
  year={2022},
  publisher={MDPI}
}

@article{jan2021design,
  title={Design and analysis of lightweight authentication protocol for securing {IoD}},
  author={Jan, Saeed Ullah and Qayum, Fawad and Khan, Habib Ullah},
  journal={{IEEE Access}},
  volume={9},
  pages={69287--69306},
  year={2021},
  publisher={IEEE}
}

@article{wang2023lightweight,
  title={Lightweight blockchain-enhanced mutual authentication protocol for {UAVs}},
  author={Wang, Weizheng and Han, Zhaoyang and Gadekallu, Thippa Reddy and Raza, Saleem and Tanveer, Jawad and Su, Chunhua},
  journal={IEEE Internet of Things Journal},
  volume={11},
  number={6},
  pages={9547--9557},
  year={2023},
  publisher={IEEE}
}

@misc{kannwischer2019pqm4,
      author = {Matthias J.  Kannwischer and Joost Rijneveld and Peter Schwabe and Ko Stoffelen},
      title = {pqm4: Testing and Benchmarking {NIST} {PQC} on {ARM} {Cortex-M4}},
      howpublished = {Cryptology {ePrint} Archive, Paper 2019/844},
      year = {2019},
      url = {https://eprint.iacr.org/2019/844}
}

\end{document}